\documentclass[reprint, prx, aps, notitlepage, nofootinbib,longbibliography,floatfix, superscriptaddress]{revtex4-1}
\usepackage{amsmath,amstext,amssymb,graphicx,bm,dcolumn,color,times, braket,caption}
\usepackage[colorlinks=true,citecolor=blue,urlcolor=red]{hyperref}
\usepackage[table]{xcolor}
\usepackage{graphicx}
\usepackage{amssymb}
\usepackage{amsmath}
\usepackage{amsthm}
\usepackage{amsfonts}
\usepackage{braket}
\usepackage{lineno}
\usepackage{subfig}

\begin{document}
\title{Global Time Distribution via Satellite-Based Sources of Entangled Photons}

\author{Stav Haldar}\email{hstav1@lsu.edu}\affiliation{Department of Physics \& Astronomy, Louisiana State University, Baton Rouge, LA 70803, USA}
\author{Ivan Agullo}\affiliation{Department of Physics \& Astronomy, Louisiana State University, Baton Rouge, LA 70803, USA}
\author{Anthony J. Brady}\affiliation{Department of Physics \& Astronomy, Louisiana State University, Baton Rouge, LA 70803, USA}\address{Current affiliation: Department of Electrical and Computer Engineering, University of Arizona, Tucson, AZ 85721, USA.}
\author{Ant{\'{\i}}a Lamas-Linares}\affiliation{Amazon Web Services (AWS), Austin, TX, USA.}
\author{W. Cyrus Proctor}\affiliation{Amazon Web Services (AWS), Austin, TX, USA.}
\author{James E. Troupe}\affiliation{Xairos Systems, Inc., Denver, CO, USA}

\pagestyle{empty}

\begin{abstract}
We propose a satellite-based scheme to perform clock synchronization between ground stations spread across the globe using quantum resources. We refer to this  as a quantum clock synchronization (QCS) network. Through detailed numerical simulations, we assess the feasibility and capabilities of a near-term implementation of this scheme. We consider a small constellation of nanosatellites equipped only with modest resources. These include quantum devices such as spontaneous parametric down  conversion (SPDC) sources, avalanche photo-detectors (APDs), and moderately stable on-board clocks such as chip scale atomic clocks (CSACs). 
In our simulations, the various performance parameters describing the hardware have been chosen such that they are either already commercially available, or require only moderate advances. We conclude that with such a scheme establishing a global network of ground based clocks synchronized to sub-nanosecond level (up to a few picoseconds) of precision, would be feasible.
Such QCS satellite constellations would form the infrastructure for a future quantum network, able to serve as a globally accessible entanglement resource. At the same time, our clock synchronization protocol, provides the sub-nanosecond level synchronization required for many quantum networking protocols, and thus, can be seen as adding an extra layer of utility to quantum technologies in the space domain designed for other purposes.
\end{abstract}

\maketitle
\section{INTRODUCTION}
\label{sec:Intro}

The ability to measure, hold and distribute time at high precision determines the limits of our scientific explorations. From a technological point of view, precise time  measurement  and  synchronization  is  an indispensable feature of communication and networking protocols, navigation and ranging, astronomical, geological and meteorological measurements, amongst others. 
The goal of this paper is to assess the feasibility and quantify the  capabilities of a concrete protocol to synchronize clocks based on the distribution of entangled photons by a constellation of satellites orbiting the Earth.  The synchronization method used is a two-way optical scheme that exploits the features of spontaneously down-converted photon pairs to provide high security time transfer.  The basics of this method were proposed and described in \cite{ho:09,AntiaTroupeQCS} and a proof-of-principle experimental demonstration  has been performed for ground based static clocks  \cite{Lee2019}, achieving synchronization precision of 51 ps in 100 s (data acquisition time) with relatively low pair rates, of order $200{\rm s}^{-1}$ for Rubidium clocks separated by up to 50 meters (also see \cite{spiess_exp} for a more recent demonstration). We call  this method Quantum Clock Synchronization (QCS) since it utilizes single-photon detection, the fundamentally random timing of the photon pair production, and for added security, the polarization entanglement between the photon pairs. 

Important questions arise when trying to extend this protocol to include satellites  in relative motion with ground stations. These questions are related to both, propagation effect---such as atmospheric losses, refraction, background counts,  beam spreading, relativistic effects etc.--- as well as questions related to network scale and connectivity ---e.g., what are farthest points on Earth that can be synced and how often the sync will occur for a concrete network. To answer these questions, we develop (building upon previous work by some of us \cite{LSU_satsim}) a software infrastructure to simulate the evolution of a satellite-network and compute the real-time quantum data communication rates between ground stations and satellite, from which we can  quantify the capabilities of the network and study its optimization. We show that such a  timing system is capable of providing higher timing accuracy than achievable with traditional Global navigation satellite system (GNSS) such as the Global positioning system (GPS) at a global scale with a modest amount of resources.

The QCS network we consider here could achieve sub-nanosecond to picosecond accuracy utilizing a constellation of nanosatellites carrying lower stability, but very low Size Weight and Power (SWaP) atomic clocks, such as the Chip Scale Atomic Clock (CSAC)\cite{Microchip}. These satellite clocks are then regularly and securely synchronized via optical links to a small group of much more stable ground-based reference atomic clocks. Such an architecture will derive a direct benefit from state of the art atomic clock technology without the requirement that these clocks be low SWaP and space qualified. In addition, the use of low cost nanosatellites would allow for cost effective upgrades of the system and robustness against hostile action against the constellation. 
Even though in the present work we use satellites as intermediaries to synchronize ground stations, we envision  global synchronization between satellites themselves and with  stable ground based clocks, forming a master clock to which interested clientele (smaller, less stable clocks) can have access to.

To put our work in perspective, we now give a brief survey of the abilities and drawbacks of classical clock synchronization techniques, in the context of providing a high-precision secure global time standard.
State of the art optical clocks can achieve a fractional frequency instability below $1 \times 10^{-18}$ \cite{opticalclocks2018} as measured by the clock's Allen Deviation (ADEV) integrated over a period of an hour \cite{oelker:04}.
Additionally, great leaps have also been made towards synchronizing such high precision, high stability clocks. Researchers have recently developed sophisticated two-way time stamping methods using radio frequency pulses that have been optimized to achieve a synchronization precision of up to tens of picoseconds \cite{rf_picosecond}. Although, these methods require a high degree of computational overhead due to the required radio frequency  propagation modeling and data processing. At the highest achievable precision, there are state of the art research results with classical optics based protocols using optical frequency combs that can provide sub-femtosecond precision time-transfer with a stability of tens of femtoseconds (over a period of several days) through a few kilometers of atmospheric turbulence\cite{Newbury2016_freq_combs, Newbury2019_quadcopter}. Such coherent techniques, however,  become extremely difficult to implement over long distances and in high loss settings \cite{Dolinar2011}. On the other hand, Time over internet protocols such as the Network Time Protocol (NTP) and the Precision Time Protocol (PTP) have global coverage but can achieve synchronization only to a few milliseconds and tenths of a microsecond, respectively, over long periods of time \cite{PTP}. More recently, the White Rabbit protocol (a refinement of PTP developed at CERN) has achieved sub-nanosecond time distribution over fiber networks over distances of hundreds of kilometers \cite{white_rabbit}. 

At the sub-microsecond level of precision, much more readily available is the use of GPS signals which can be used to provide almost continuous synchronization with an error of tens of nanoseconds ($<40$ ns, 95\% of the time)\cite{Zhu_GPS_relativity}. While originally developed by the US Department of Defense for precision global navigation and positioning to support military applications, GPS has become a defacto global time standard. In fact, a  study in 2017 stated that of the estimated \$1.4 trillion of economic benefits that GPS has generated since it was made available for civilian and commercial use in the 1980s, well over half are directly based on precision timing and synchronization \cite{RTI}. The ubiquity of GPS time information has been an important factor in the explosion of applications and technologies that rely on such a global time standard, e.g. 5G telecommunication networks. However, there are well known issues regarding the security of GPS, e.g the relative ease of jamming and spoofing GPS timing signals. In addition, while the nominal performance of GPS is impressive, future technologies such as 6G+ communication and quantum networks will require clock synchronization better than provided by GPS or other Global Navigation Satellite Systems (GNSS) \cite{6Gplus,GE_2021}. 

Beyond the performance of GPS and other GNSS, another drawback is the high cost of these systems. The satellites required are very costly, with their large size required so as to contain the highly stable atomic clocks needed to achieve the required degree of synchronization. GPS satellites are regularly synchronized to each other by updates from ground stations. However, in order to maintain the system's specified time accuracy with sufficient holdover to bridge these updates, the GPS satellite atomic clocks require stability on the order of $10^{-14}$ to $10^{-15}$ in terms of frequency accuracy \cite{GNSS_AtomicClocks}. Given these constraints in this paper we propose a QCS network with the purpose of complementing existing classical techniques such as the GPS in providing a more precise, robust and secure global time standard.

This paper is organized as follows. 
Section \ref{sec:QCS_protocol} describes the protocol used to synchronise satellite and ground station clocks and to determine the clock offset  from the two-way time-stamp correlation functions. Section \ref{sec:Sat_based_QCS} discusses the necessity and advantages of using a satellite constellation for building a global QCS network. Section \ref{sec:simulations} gives a detailed account of our numerical simulations to estimate the performance of QCS networks and also provides a concrete example in the form of the QCS network servicing the continental US at a sub-nanosecond time sync precision. In Section \ref{sec:main_results} we summarize the key takeaways from our simulations. We then move on to Section \ref{sec:Conclusions} that collects the main conclusions, discusses some shortcomings of the techniques used in this work and also proposes directions for future explorations.

\section{Quantum Clock Synchronization Protocol}\label{sec:QCS_protocol}
This work is primarily related to the question of time transfer or estimating the \emph{time offset} between two separate clocks. We use the term Quantum Clock Synchronization (QCS) to describe the process of the exchange of quantum signals to estimate the time offset between remote clocks (see \cite{Jozsa:2010,giovannetti:01,giovannetti:04,valencia:04,PhysRevA.65.052317,ho:09,Jon_QCS_2018} for pioneering ideas on this field). The QCS protocol is a type of Optical Two Way Time and Frequency Transfer (O-TWTFT) scheme. Recall that a general O-TWTFT method uses signals transmitted symmetrically in both directions between Alice's and Bob's clocks to estimate the relative clock offset and frequency difference (sometimes referred to as frequency skew) between them. The central component of the clock offset estimation is the calculation of the cross-correlations between the local timestamps of the photons produced by Alice and  those received by Bob, and vice versa, out of which one can obtain the clock offset, as we explain below.

For classical TWTFT methods, the signals are pseudo-random codes encoded into an optical or radio frequency carrier. The broad frequency spectrum of these pseudo-random signals will maximize the precision of the clock offset due to the corresponding sharp cross correlation peak. In the case of a relative frequency difference between the two clocks, this method can be extended to also estimate the frequency difference by estimating the change in the clock offsets as a function of the local time of one of the clocks. 

In the case of the QCS protocol, the classical optical signals are replaced with pairs of individual photons created via spontaneous down-conversion (SPDC) by pumping a nonlinear optical crystal. Due to conservation of energy, the time of birth of the photons are very highly correlated with each other, typically on the order of 10-100 femtoseconds \cite{spdc_time_corr1970, Shih2004}. In addition, the SPDC pair production process is quantum mechanically random with photon pair production following a Poisson distribution. The photon pair production time is itself used as the (now truly) random ``code'' to be shared between Alice and Bob. Alice and Bob each have a SPDC source and locally detect one photon of each pair produced, recording a timestamp of each detection. The other photon from each down-converted pair is then transmitted to the other party and its detection is timestamped. The clock offset is then estimated by the difference of the two cross-correlation peaks as with the classical O-TWTFT scheme. 

\subsection{Clock Offset Estimate}\label{subsec:extraction}

Here we review the QCS protocol in detail, as first reported in \cite{AntiaTroupeQCS} and with the first experimental demonstration by Lee, \emph{et al.} \cite{Lee2019}. Following Ho, \emph{et al.} \cite{ho:09} we denote the numbers measured by Alice's (Bob's) local clock by $t$ ($t'$) with a subscript denoting a particular indexed event. If Alice and Bob were at the same spatial location detecting the same pair event, the difference between the times of detection as measured by their local clocks would be $\delta=t-t'$, and this $\delta$ would be the time offset that we aim to determine. If Alice and Bob are at separate locations, the time of propagation of a signal between Alice and Bob is denoted as $\Delta t_{AB}$ ($\Delta t_{BA}$ for propagation in the opposite direction). The round trip time of a signal originating from either Alice or Bob is $\Delta T=\Delta t_{AB}+\Delta t_{BA}$.

   \begin{figure*} [t]
   \begin{center}
   \begin{tabular}{c} 
   \includegraphics[width=\textwidth]{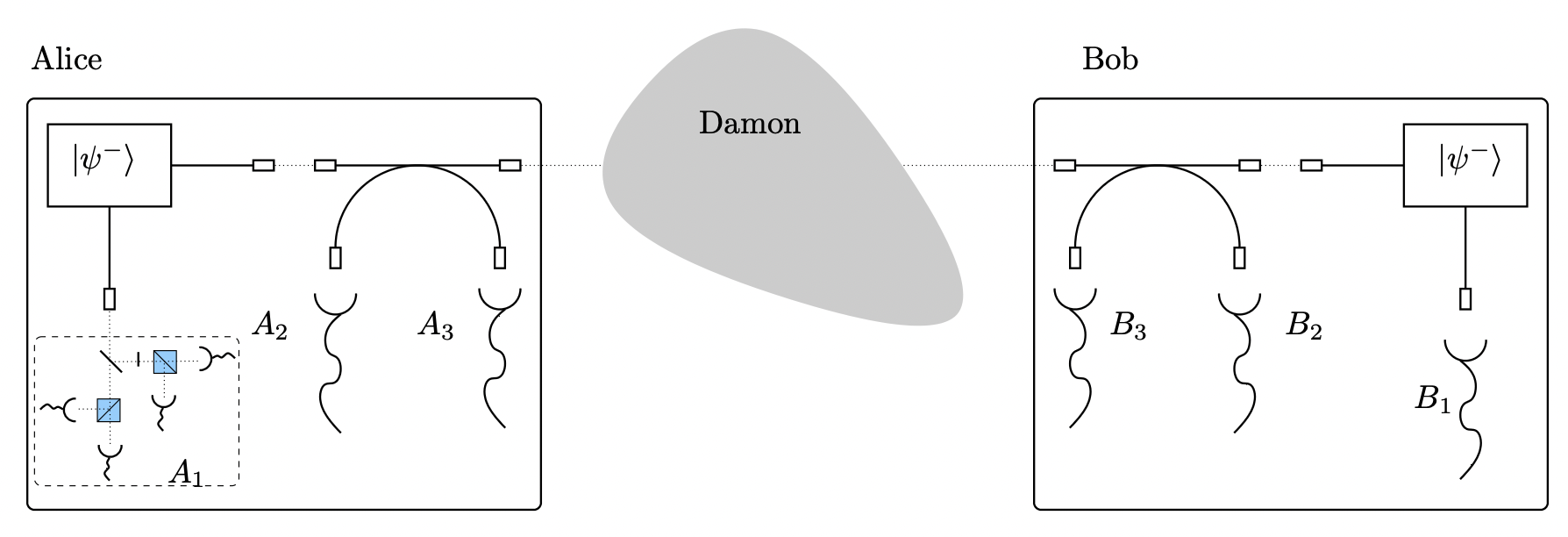}
   \end{tabular}
   \end{center}
   \caption[example] 
   { \label{fig:example} 
Schematic for experimental implementation of the QCS protocol using entangled photons (details of this example setup are not important for the following discussion and are included for completeness). Alice and Bob each have a source of polarization entangled pairs ($\ket{\psi^{-}}$) produced by spontaneous parametric down-conversion (SPDC) and a set of single photon detectors within their secure lab (denoted by a solid line). Each mode of propagation of the photons ends in a detector cluster able to perform polarization measurements, but only the cluster labeled as $A_1$ is fully represented in the figure. One member of the SPDC pair is detected locally at detector cluster $A_1$ on Alice's side (cluster $B_1$ on Bob's side). The other member of the pair is sent into a single mode fiber and propagated through a channel controlled by an adversary, Damon. Each of the propagating photons has a chance of being detected on the remote side by $B_2$ ($A_2$) for pairs originating at Alice's (Bob's) side. Times of arrival for all detected photons are recorded in each lab with respect to a local clock. Detectors $A_3$ and $B_3$ are under the control of either Alice or Bob and are included for completeness but do not play a part in the discussion. The detector cluster illustrated for $A_1$ represents a possible passive measurement scheme for a CHSH inequality. It uses a beam splitter followed by two polarizing beam splitters oriented at the appropriate angles for projection into the desired polarization state.}
   \end{figure*}

Additionally, since the channel is a single spatial mode and the signals propagating between Alice and Bob are identical in all degrees of freedom apart from propagation direction, the protocol is based on the assumption $\Delta t_{AB}=\Delta t_{BA}=\Delta t$, which we will refer to as channel reciprocity and further discuss it below.  
To calculate the absolute time difference between clocks, $\delta$,  consider a photon pair produced at Alice's site. One of the members of the pair is detected locally at detector\footnote{Alice has two detectors, $A_1$ and $A_2$, one dedicated to measure photons produced locally by Alice, and the other to detect photons received from Bob. The same is true for Bob.} $A_1$ and the other member of the pair travels to Bob accumulating a travel time $\Delta t_{AB}$ and getting detected at $B_2$. For any particular photon pair event produced at Alice's site, the difference between the time labels recorded at Alice and Bob will be:
\begin{equation}
    t'-t=\Delta t_{AB}+\delta. \nonumber
\end{equation}
Similarly for any photon pair produced at Bob's site:
\begin{equation}
    t-t'=\Delta t_{BA}-\delta. \nonumber
\end{equation}
Therefore, the information about the offset is encoded in the time stamp differences. However, in practice to extract the offset, one finds the added difficulty that Alice and Bob both receive many photons, either from the other party or from ambient noise, and we need a way to identify which photons belong to the same pair. This is where correlation functions become useful. The differences between the time labels can be extracted by calculating a cross-correlation between events at both sides. Consider first events produced at Alice's site. The detection events are translated into a distribution as:
\begin{eqnarray}
a(t)&=&\sum_i \delta\left (t-t_i\right) \mathrm{d}t \nonumber \\
b(t')&=&\sum_j \delta\left (t'-t'_j\right) \mathrm{d}t', \nonumber
\end{eqnarray}
where $i$ and $j$ index arbitrary detection events which can arise either from the production of entangled pairs of photons or from other detector triggers such as stray light, dark counts, etc.
The cross-correlation is computed as
\begin{equation}
c_{AB}(\tau)=(a\star b)(\tau)=\int a(t)b(t+\tau)\mathrm{d}t, \nonumber
\end{equation}
and, for sufficiently high signal-to-noise ratio,  will have a maximum at $\tau=\tau_{AB}=\Delta t_{AB}+\delta$. Likewise, if we consider those pairs created on Bob's site, we can extract another cross-correlation,
\begin{equation}
c_{BA}(\tau)=(b\star a)(\tau)=\int b(t)a(t+\tau)\mathrm{d}t, \nonumber
\end{equation}
which will have a maximum at $\tau=\tau_{BA}=\Delta t_{BA}-\delta$.
From these we can extract both the round trip time ($\Delta T$) and the absolute time difference between clocks without making any prior assumptions about the length of the path between Alice and Bob:
\begin{eqnarray}
\Delta T &=& \tau_{AB}+\tau_{BA} \nonumber \\
\delta &=& \frac{1}{2}\left (\tau_{AB}-\tau_{BA}\right ). \nonumber
\end{eqnarray}
From this we can see that the accuracy of the clock offset estimate is determined by the accuracy of estimates for each of the cross-correlation peaks. 

In the discussion above we have assumed a static situation with no motion between Alice and Bob. In the case of inertial relative motion the effect will be to spread out the correlation function by an amount proportional to the time over which the estimate is made, i.e., over the acquisition time. This acts essentially as an additional effective clock drift between Alice's and Bob's clocks. The length of the required acquisition time, the timestamp resolution, and the relative velocity, will determine the effect that the motion has on the cross-correlation peak. If the relative velocity is known to within a maximum error, then the timestamp data can be corrected to compensate for this ``stretching" effect with some residual uncertainty. For example, if the relative velocity is known to a maximum error of 1 cm/s, for an acquisition time of 250 milliseconds, the residual uncertainty in the correlation peaks will be less than 10 picoseconds (assuming that for short enough acquisition times the relative velocity remains constant). On the other hand if the relative velocity is not known to high precision, a parameter search can be performed over the timestamp data to estimate both the velocity and the constant clock offset using methods already indicated in \cite{ho:09}.

Note that the quantum entanglement of photon pairs created at Alice’s and/or Bob’s locations does not play a direct role in determining the clock offset estimate, except that the production of entangled pairs by SPDC ensures that the photons in each pair are generated within a time window typically of a few 100 fs – several orders of magnitude smaller than the time scales involved in the synchronization protocol. However, the entanglement between photon pairs plays a crucial role in increasing the security of this protocol to malicious attacks since it can be used to directly verify the detected photons used for the clock offset estimates. Violation of Bell's inequality implies that the correlations in the photon pair's polarization could not have been fully copied and therefore spoofed by an adversary. Under the assumption of channel reciprocity (i.e. that the propagation time is the same in each direction), this verification ensures the security of the estimated clock offset. This follows due to the bidirectional nature of the two-way QCS protocol since reciprocity implies that symmetric delays introduced into the channel by an adversary will have no effect on the estimated clock offset. In the case of free space optical channels through the atmosphere, only very small deviation from full reciprocity are typically incurred.  For instance, between a Medium Earth Orbiting (MEO) satellite and ground station the maximum error due to turbulence induced non-reciprocity is predicted to be less than 10 femtoseconds \cite{Belmonte2017, Taylor2020}. Partial reciprocity and other non-idealities of practical implementations would likely place a limit on the achievable \emph{secure} accuracy of the clock synchronization protocol. An analysis of the system requirements for using the detected entangled photons to warrant a specified level of security for the distributed time information is the focus of future work.


\section{Satellite-Based QCS}\label{sec:Sat_based_QCS}

\begin{figure} [ht]
   \begin{center}
   \begin{tabular}{c} 
   \includegraphics[height=6cm]{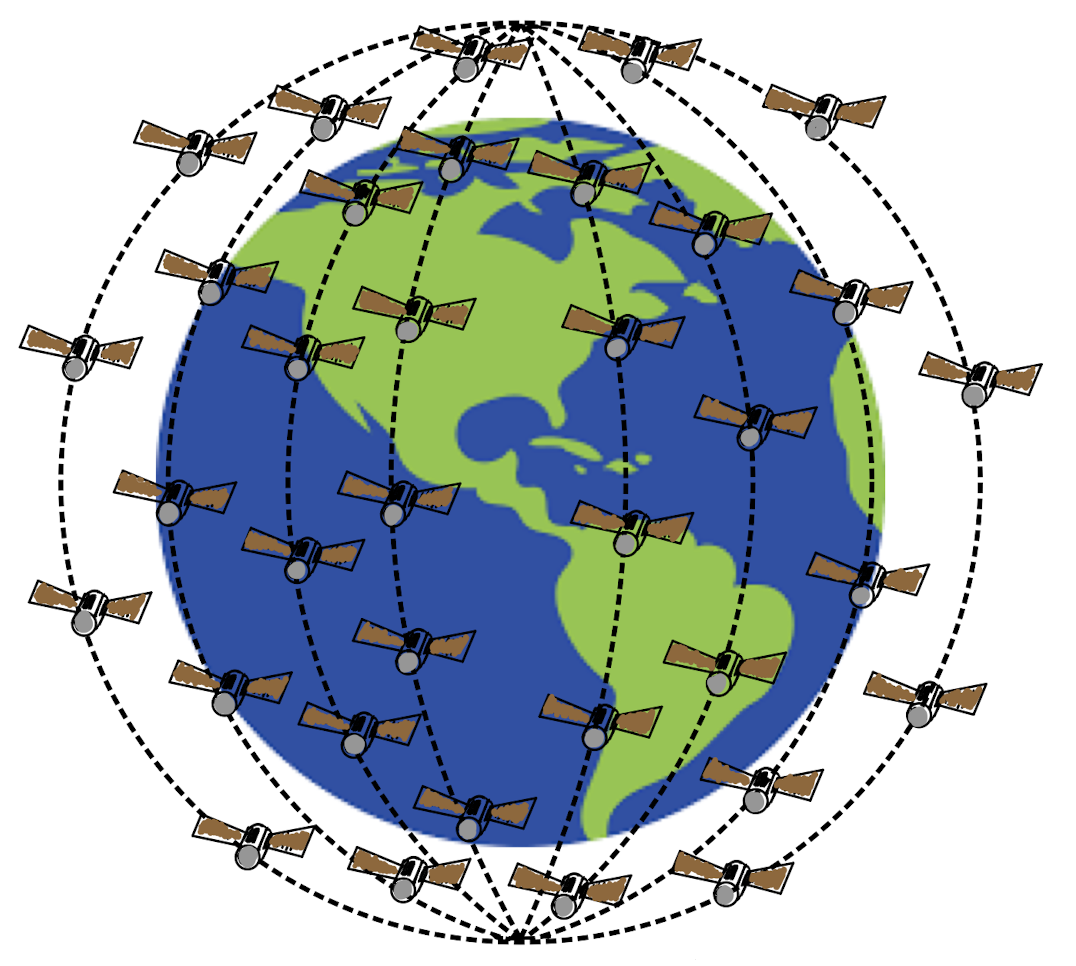}
   \end{tabular}
   \end{center}
   \caption[example] 
   {\label{fig:sat_constellation}
A constellation of satellites used for global time distribution. Each satellite is equipped with an entangled photon-pair source along with photodetectors and transceivers (figure adapted from \cite{LSU_satsim}). The satellites transmit and receive quantum signals to and from ground stations through a bi-directional quantum communication channel (downlink and uplink channel), which is used to synchronize onboard satellite clocks with terrestrial clocks using the QCS protocol mentioned in Section \ref{sec:QCS_protocol}.}
   \end{figure}

The technical requirements of QCS, while not as stringent as most quantum communication tasks, does share many of the same features, e.g. the requirement of a sufficiently high entangled bit (ebit) rate between communicating parties. A significant amount of current research and development is focused on the use of satellites to go beyond the limitations of terrestrial fiber-based quantum communication networks. For example, the quantum repeater-less fiber optic based secret key rate bound is surpassed beyond 215 km for a satellite at altitude of 530 km \cite{Pan_exp_crypto, Sidu2015}. This means for distances beyond a few hundred kilometers between commmunicating parties, and in the absence of quantum repeaters, free-space communication via satellites will provide higher ebit rates. A key role in developing feasible long distance quantum communication implementations has been played by the transition to hybrid space-terrestrial quantum communication network architectures combining satellites and ground stations equipped with optical telescopes with metropolitan scale fiber optic networks (see Figure \ref{fig:sat_constellation}). This is because longer distance realizations of fully terrestrial quantum networks are hindered by the exponential losses associated with ground-based communication channels (primarily fiber optic cables)\cite{LSU_satsim}. Unlike the classical information encoded in classical optical signals, the quantum information encoded via quantum communication protocols cannot be amplified due to fundamental limits on copying quantum information. This places fundamental limits on directly transferring quantum information through lossy channels. Large number of high fidelity quantum repeaters and/or quantum memories could improve the situation to some extent, but their current performance levels are below those needed for mature applications \cite{Sidu2015}, and furthermore, it would very likely be impractical to place such devices in difficult terrain, e.g. mountains or oceans.

Recent proof of principle experiments through the Micius satellite have indeed shown the effectiveness of a satellite-based quantum communication channel for large scale quantum networks \cite{Pan_exp_crypto, Pan_exp_teleport, Pan_satellite_QCS}. We quote a most striking observation from this seminal work to point out the inevitability of shifting to satellite based platforms: ``As a comparison, using the same four photon source and sending the teleported photon through a 1,200 km telecommunication fibre with loss of 0.2 dB $km^{-1}$, it would take 380 billion years (20 times the lifetime of the Universe) to witness one event, assuming the detectors have zero dark counts'' \cite{Pan_exp_teleport} (It is assumed that quantum repeaters are not used. Progress has been made towards implementing quantum repeaters but the technology is far from mature \cite{Sidu2015}). Other groups are also currently working to use small satellites to perform basic quantum communication tasks such as Quantum Key Distribution (QKD) \cite{Nano_sat,Sidu2015}. 

\begin{figure} [ht]
   \begin{center}
   \begin{tabular}{c} 
   \includegraphics[height=5cm]{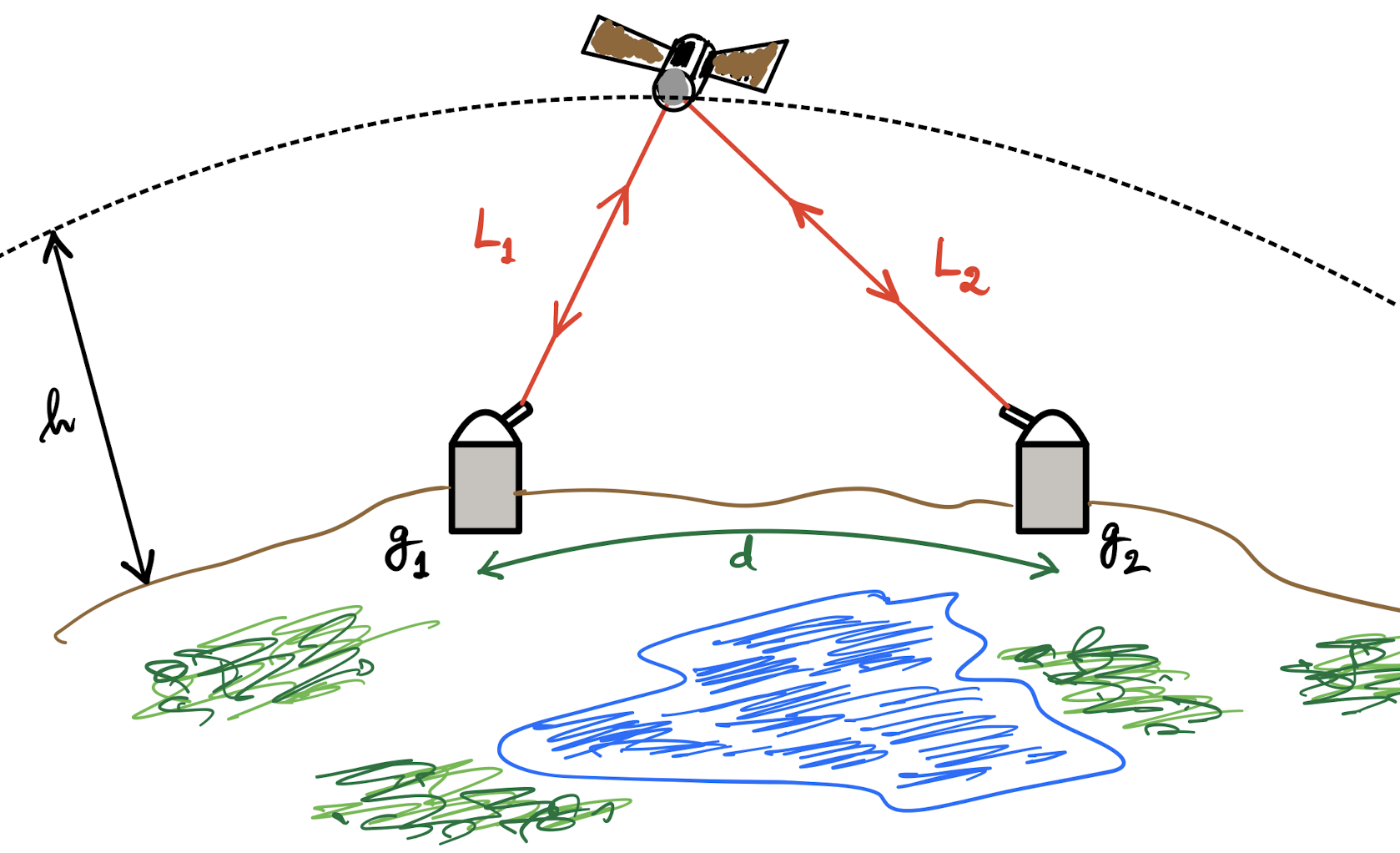}
   \end{tabular}
   \end{center}
   \caption[example] 
   {\label{fig:duallink} 
Dual-link depiction for simultaneous clock synchronization between two ground stations ($g_1$ and $g_2$) with geodesic distance $d$ between them (Adapted from \cite{LSU_satsim}). A LEO satellite at altitude $h$ with asymmetric link distances $L1$ and $L2$ to the two ground stations.}
   \end{figure}

Consider now that Alice and Bob are separated in a way that makes it inefficient to exchange photons directly between the two parties.  A quantum network between different cities is an example of such a scenario.  The distances are large enough ($\approx$ 1000 km) to make direct communication through standard optical fiber channels (even with repeaters) less efficient and resource consuming than communication through a network of intermediary satellites in low Earth orbits. The satellites are to be used as intermediaries in the sense that ground station A can be synced to a satellite and then the same satellite could be synced to the ground station B. This can either happen simultaneously as shown in Figure \ref{fig:duallink}, or if the ground stations are too far apart, the two clock offsets can be sequentially estimated and compared after the satellite passes within range of both. In the latter case the maximum time allowed between sync at stations A and B is determined by the stability of the satellite clock and is quantified by the hold over time $\tau$. If all three clocks involved are relatively stable within the time this protocol is executed then the clocks at A and B can be successfully synchronized in either case. This time includes the acquisition times ($T_a$) needed for individual sync events at A and B plus the hold over time $\tau$. Thus, the elementary task of this protocol is to synchronize a ground station and a satellite. In other words, we do not consider inter-satellite communication, and therefore the sync between two ground stations must be established through a common satellite. This situation is of interest as a low-resource  way of synchronizing users that span up to a continent sized geographical area---e.g.\ a handful of select cities across the contiguous US. 

\section{Simulation results} \label{sec:simulations}
The analysis of a QCS network as described in Sections \ref{sec:Intro} and \ref{sec:Sat_based_QCS} is a complex task given that there are a number of variables involved. At the level of the fundamental QCS link between a satellite and ground station, one must first analyse the role of hardware parameters such as receiving and transmitting telescope radii, detector efficiencies, clock stability and source rates, etc. At the same time, one must also model a dynamic quantum communication link, since the satellite-ground station distance is constantly changing (also a satellite is not always visible from a given ground station), effectively changing the transmissivity of the channel. Further, at the level of the network, the number of design parameters such as number of satellites, orbit selection (altitude, inclination etc.) is also large. This complexity makes this problem suitable to be computer simulated and we use this route to model the dynamical quantum communication link and to analyze quantitatively  the performance of various network configurations/designs. As we discuss in detail in the next section, our code has the ability to simulate different scenarios in terms of resource constraints. The code is written in Python and uses parallelization in order to simulate large network sizes with tens of satellites and several cities across the globe.

Subsection \ref{subsec:dynamic_simulations} describes our simulations to characterize the performance of the QCS protocol for the purpose of synchronizing ground stations via small LEO (also MEO) satellites. We will  simulate the real time motion of satellites relative to ground stations and obtain the periods of time along the day over which synchronization is possible. In order to quantify the capabilities of the network, rather than computing the cross-correlation functions described in section \ref{subsec:extraction}, we will use quantum data communication rates (measured as entangled bits (ebits) shared per second)  between satellites and ground stations as a proxy for sync precision. We will require a minimum ebit threshold for the sync to be able to occur at all.
The use of ebit rates is highly convenient, since it provides an efficient tool to compute coverage area for time distribution from each satellite. This allows us to perform the simulations for large number of network nodes (ground stations) and over long time scales, since the cross-correlation functions need not be simulated photon by photon. 
Nonetheless, we will justify a posteriori the use of ebit rates as proxy for sync precision through ``static simulations'' in section \ref{subsec:static_simulations}. 

\subsection{Dynamic simulation}
\label{subsec:dynamic_simulations}

This scenario takes into account the motion of satellites and ground stations around the globe, in order to evaluate the periods of time along the day that ground stations are ``in view'' of  satellites. This is done at the level of quantum data communication rates (ebits shared per second) without looking at the actual correlation functions. As summarized above,  ebit rates are used as a proxy for the level of achievable precision, using the intuition that larger ebit rates correspond to higher correlation function peaks and hence higher precision. This intuition is backed up by static simulations in subsection  \ref{subsec:extraction},  where the actual correlation functions are simulated. As we show ahead, our simulations provide useful estimates for the capabilities of the network, in terms of the quantity  of resources (number of satellites,  number of orbits, etc.) needed to perform specific sync tasks. The results from Table \ref{tab:static_effect_no_jitter_50ps_res} in section \ref{subsec:static_simulations} indicate that for the range of link losses encountered for a LEO orbit (500 km altitude) the protocol is always successful if the ebit rate is greater than approximately 200 ebits/s. From these observations, it follows that, if the sync is considered successful only if the clock offset can be evaluated better than (or equal to) 1 ns precision, then setting the cut off for the quantum communication rate between the ground stations and satellites at 200 ebits/s ensures success of the protocol. The small acquisition time requirement (250 ms) also justifies the use of ebit \emph{rates} as a proxy for sync precision.

\subsubsection{Details of the simulation technique}
\label{subsubsec:sim_details} 
Our first goal is to apply the protocol described above to synchronize a satellite (Alice) and a single ground station (Bob) via a bi-directional quantum communication channel, i.e.\  a downlink and an uplink, and then use it to synchronize two ground stations that come in view of the same satellite. 

We need to address two sets of questions: (i) characterization of the quantum communication channel, which will ultimately limit quantum data rates, and (ii) the dynamics of orbiting satellites. First, we look at the quantum communication channel between a ground station and the satellite. For concreteness, let us focus on a downlink channel; similar results hold for the uplink.  We consider only a lossy channel assuming clear skies and ignoring any background noise from spurious sources ---this can be incorporated later by introducing signal-to-noise  thresholds or by choosing a higher cut-off rate \footnote{While estimating the cut-off rate using static simulations in Section \ref{subsec:static_simulations} a dark count rate of 1000 Hz was assumed, hence the cut-off rate $\mathcal{R}_c = 200$  ebits/s includes the effect of noise due to dark counts. Similarly, the effect of stray light can be included by raising the cut-off further.}. Photons are either transmitted through the channel or lost in transmission. We characterize various loss mechanisms by their transmittance values $\eta$, which is the fraction of the received optical power to the transmitted power. The dominant sources of loss are (i) beam spreading (free-space diffraction loss, pointing error etc.),  $\eta^{\rm dwn}_{\rm fs}(L, h)$, (ii) atmospheric absorption/scattering,  $\eta^{\rm dwn}_{\rm atm}(L, h)$, and (iii) non-ideal photodetectors on the satellite and on the ground, with efficiencies given by  $\kappa_{\rm sat}$ and $\kappa_{\rm grd}$ respectively.
The superscripts refer to the downlink. The transmittances are functions of the link distance $L$ (physical distance) between satellite and receiver and $h$ which is the satellite altitude. Simple, analytic formulae are used to estimate $\eta^{\rm dwn}_{\rm fs}$ and $\eta^{\rm dwn}_{\rm atm}$, following Ref.~\cite{LSU_satsim}. Then, given an onboard source which generates entangled photons at an average rate of $\mathcal{R}$ ebits per second, we can estimate the quantum data communication rate (ebit rate) between a satellite and a ground station for the downlink by

$$\mathcal{R}^{\rm dwn}=\mathcal{R}\, \eta^{\rm dwn}_{\rm fs}\, \eta^{\rm dwn}_{\rm atm}\, \kappa_{\rm sat}\, \kappa_{\rm grd}\, .$$

The precision to which Alice and Bob can synchronize their clocks depends on the amount of successfully detected, correlated photons transmitted through the bi-directional communication channel over some acquisition period. Hence, the quantum data rate pair
$Q:=\Big(\mathcal{R}^{\rm dwn},\mathcal{R}^{\rm up}\Big)$ 
serves as a useful performance metric for the clock synchronization protocol. We generalize this to a network of satellites and ground stations by indexing the quantum data rate pair as $Q_{ij}$, where the first index corresponds to the i-th satellite and the second index corresponds to the j-th ground station.

Next, we incorporate dynamics into the communication channel. Since the satellites are in motion with respect to the ground stations, the quantum data rates will generically change as a function of time, since the transmittance values are a function of the  physical distance $L$ (and visibility) between the satellite and ground station. For simplicity, we assume circular orbits for all satellites. We adapt and extend software previously created by authors in Ref. \cite{LSU_satsim}, to simulate satellite and ground station motion and compute physical distances between them as functions of time. These distances are then used to evaluate transmittances and the quantum data rates $Q_{ij}$. 

In our simulations, the parameters describing the hardware design have been reasonably chosen based on recent demonstrations \cite{Nano_sat,Pan_exp_crypto}, and are  representative of the current state-of-the-art. The operating wavelength of the sources was chosen to be 810 nm, while the source strength was set to 10 million pairs per second. The detectors onboard the spacecraft are assumed to be non-cryogenic, passively quenched, Geiger-mode avalanche photodiodes (GM-APDs) that are 45\% efficient at a wavelength of 800 nm. For this design study, detector efficiencies were set to 50\% as a representative value for a broad class of APDs. Lastly, the apertures for the satellite and ground station telescopes were chosen to be of 10 cm (fill-factor of 80\%) and 60 cm respectively, which represent typical optical communications sizes that are used for classical LEO laser communications \cite{planewave:2020,cubesat:2020,Oi2017}. Apart from these fixed hardware parameters, we also have several other variable parameters such as orbit altitude, number of satellites, distribution of satellites amongst orbits, orbit inclination, stability of satellite clocks (quantified by the hold over time $\tau$), and cut off rates $\mathcal{R}_c$.

\begin{figure}
   \centering
   \begin{tabular}{c} 
   \includegraphics[height=8cm]{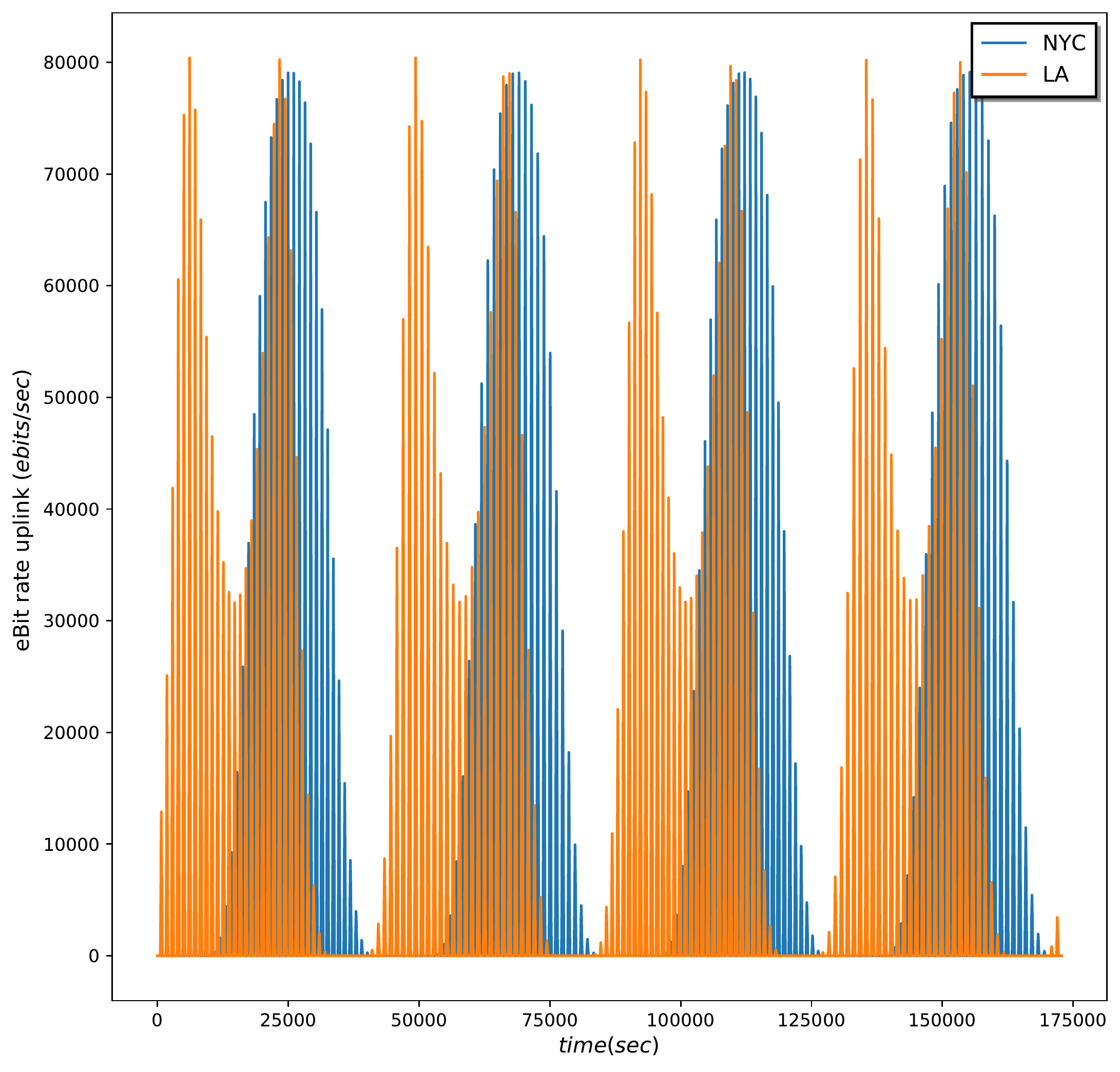}
   \end{tabular}
\caption
   { \label{fig:NYC_LA_uplinks}
Connection traces (uplink ebit rates) for two cities in the US: New York City and Los Angeles. The satellite network is comprised of 10 satellites --- two tilted polar orbits ($50^\circ$ and $-50^\circ$ to the Earth’s axis of rotation) of 5 satellites each in a 500 km LEO. Connection is considered established when the ebit rate exchanged between the satellite and ground station is greater than the cut off rate of 200 ebits/s. Each sharp vertical line corresponds to a single satellite pass (which would be a broader curve if we zoomed in to a smaller time scale), and consecutive vertical lines correspond to passes of consecutive satellites within an orbit. The bunching of vertical lines is due to the fact that there are many satellites in each orbit. The recurrence of this bunch occurs after a 6 hour period (approximately 20,000 seconds after the first bunch in the trace), when an orbit comes in view of the cities. The separation between the peaks for the two cities indicates the fact that satellites pass over these cities at different times.}
\end{figure}

We study the effect that a change in these parameters has on the performance of the network, which we assess through a few simple figures of merit defined in the following sections.
The fundamental output of our simulations is the {\em connection trace} between satellites and individual ground stations, which is the quantum data communication rate for uplinks (which is always weaker than the downlink, primarily because of the smaller receiving telescope onboard the satellite) as a function of time. 
For example, Figure \ref{fig:NYC_LA_uplinks} shows the connection trace for uplinks from New York and Los Angeles to a constellation of 10 satellites distributed equally into two tilted polar orbits. 

The connection traces provide information about the number of ebit pairs available at any given time that could be utilised to synchronize a ground station with a satellite. Nonetheless, all these ebits are not useful for the synchronisation between two ground stations. In order to estimate the sync quality between two ground stations we must only consider ebits which are shared between two ground stations and a common satellite. To that end, from the overlap between connection traces of two cities we can compute {\em sync traces}. This overlap can be either instantaneous or over a time window $\tau$, where $\tau$ is equal to the hold over time of the clock onboard the satellite. More concretely, the sync trace for a city is different from zero at a given time only if the other city also receives an ebit rate above the threshold from a common satellite. This can happen if a common satellite has been in view of both the cities within an time interval $\tau$. If $Q_{1j}$ and $Q_{2j}$ are the uplink ebit rates from ground stations 1 and 2 to some common satellite (index j), then the sync trace $Q_1(t)$ for city 1 is given by (also similarly defined for city 2):

$$Q_1(t) = \max\limits_{j}{Q_{1j}(t)}$$ where,
\begin{widetext}
\begin{equation*}
    Q_{1j}(t) =  \begin{cases} 
      Q_{1j}(t), & {\rm if} \; Q_{1j}(t)>\mathcal{R}_c \; \; {\rm and} \max\limits_{\{t-\tau < t' < t+\tau\}}\{Q_{2j}(t')\} \,  \, > \mathcal{R}_c \\
      0, & {\rm otherwise}
   \end{cases}
\end{equation*}
\end{widetext}

\begin{figure}
   \begin{center}
   \begin{tabular}{c} 
   \includegraphics[height=8cm]{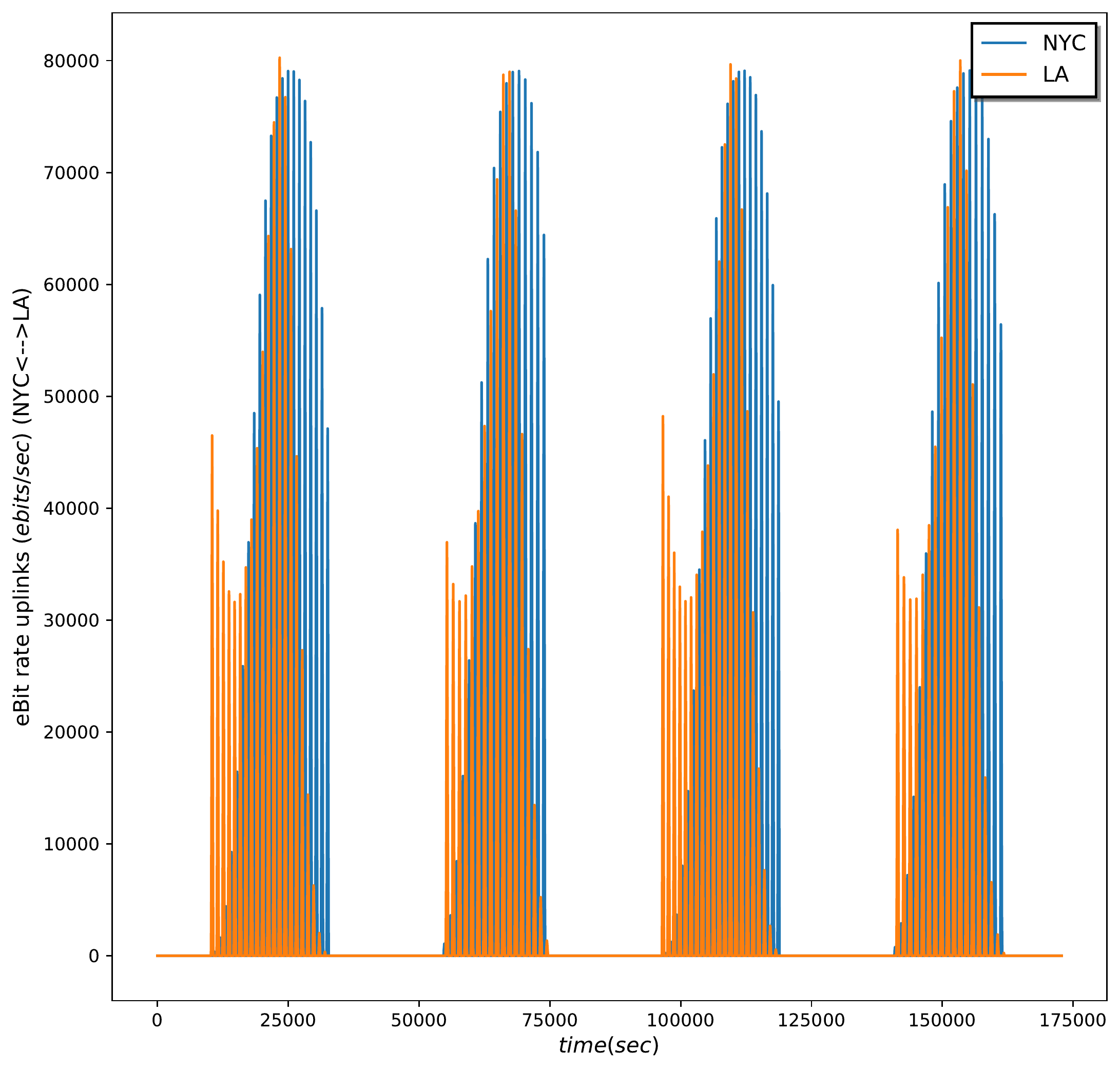}
   \end{tabular}
   \end{center}
\caption
   { \label{fig:NYC_LA_sync_uplinks}
Sync traces for a pair of cities in the US: New York City--Los Angeles. The satellite network is comprised of 10 satellites --- two tilted polar orbits ($50^\circ$ and $-50^\circ$ to the Earth’s axis of rotation) of 5 satellites in a 500 km LEO. The sync trace of a city is non-zero only when it is in view of a satellite (with ebit rate above the threshold), and 
the other city from the pair has also been (or will have also been) in view of the same satellite within an interval equal to the hold over time of the clock onboard the satellite. For this figure $\tau = 600$ s. Sync traces are chopped up versions of the connection traces, indicating regions of simultaneous connection (within a $\tau$ window). Compare with Figure \ref{fig:NYC_LA_uplinks}.}
\end{figure}

\begin{figure*}
\begin{center}
    \subfloat[Hold over time $\tau = 0$ s.]{%
      \frame{\includegraphics[width=0.4035\textwidth]{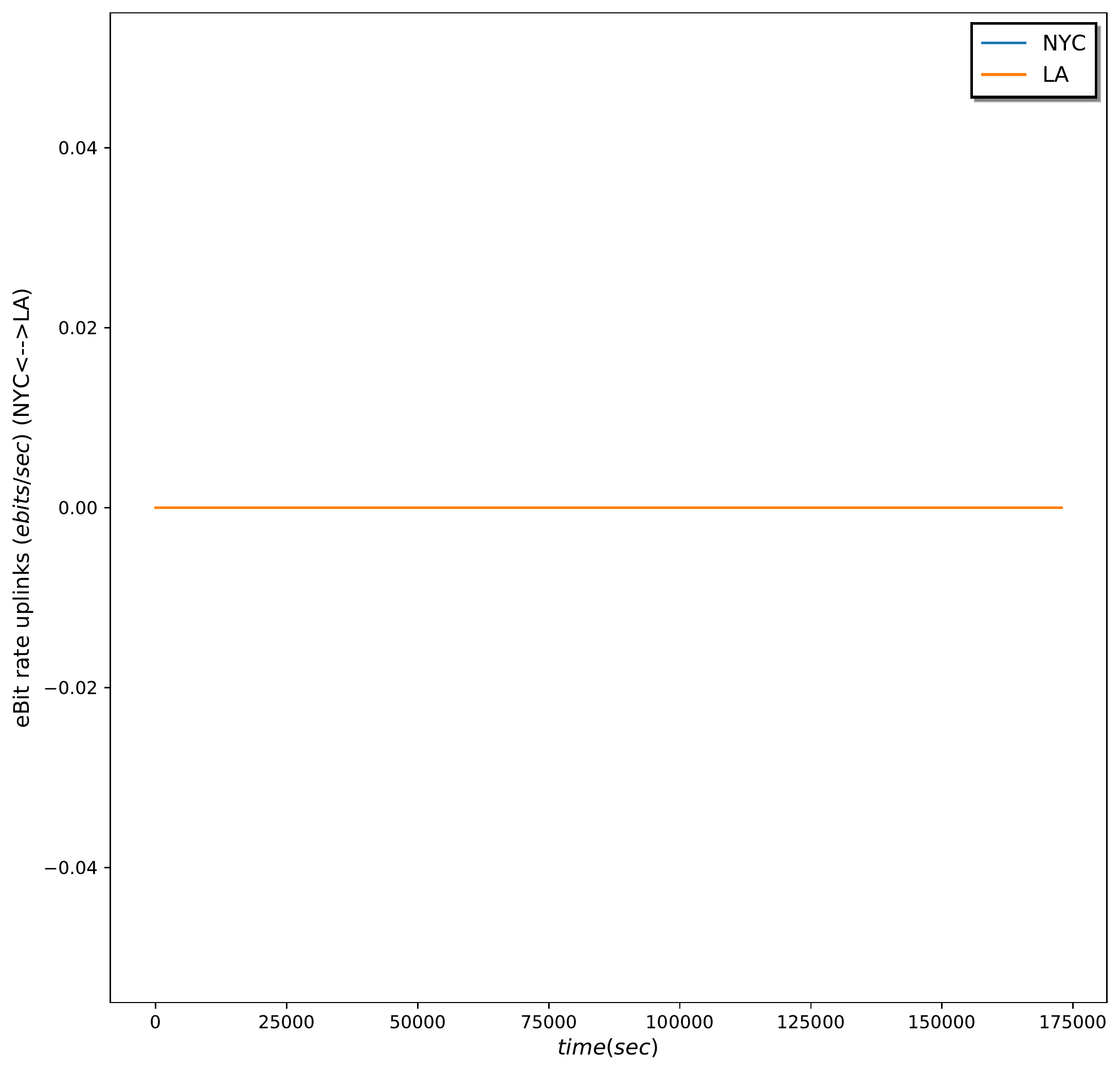}}%
      \label{fig:tau0}%
    }
    \subfloat[Hold over time $\tau = 100$ s.]{%
      \frame{\includegraphics[width=0.4\textwidth]{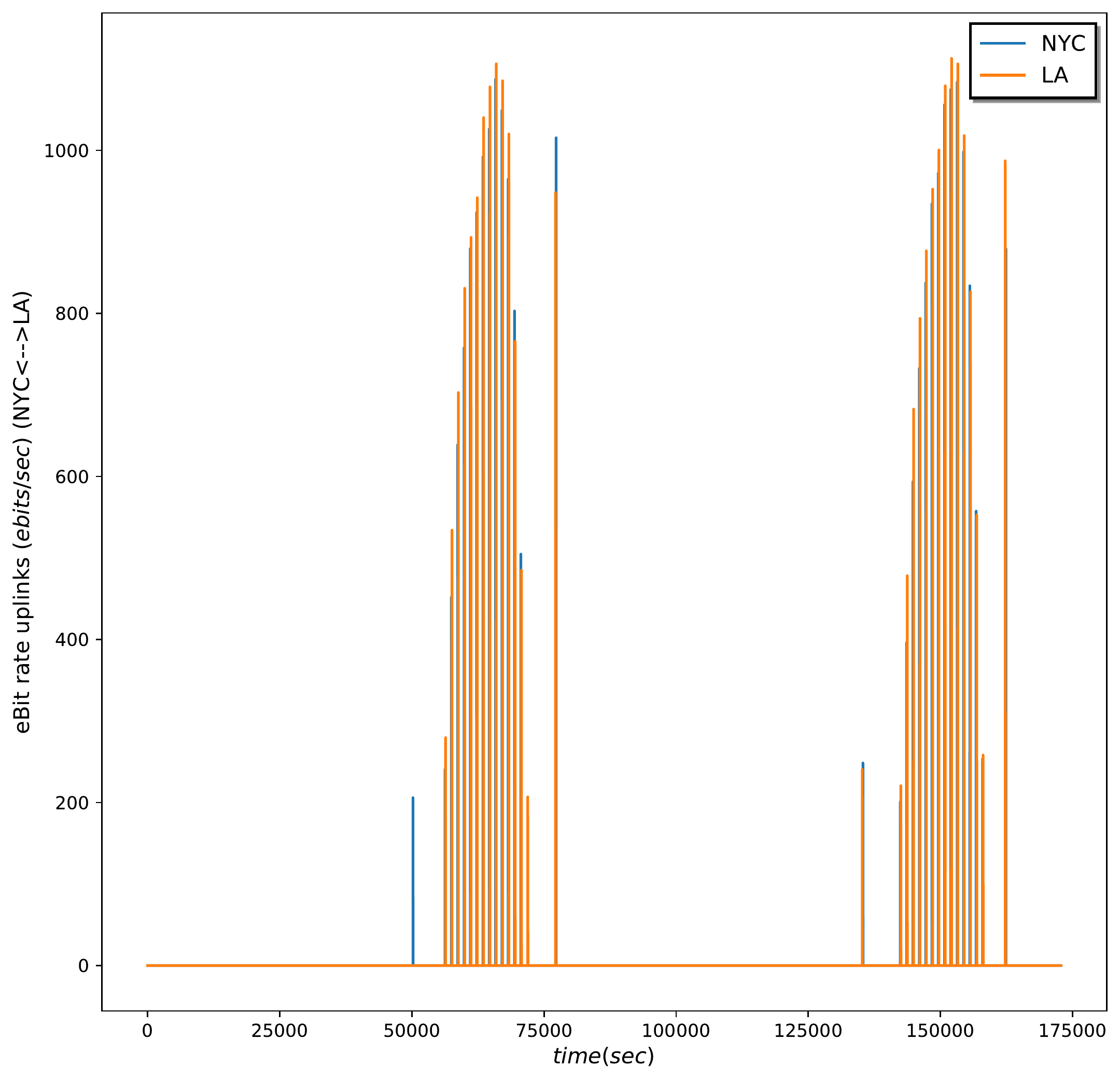}}%
      \label{fig:tau100}%
    }
    
    \subfloat[Hold over time $\tau = 200$ s.]{%
      \frame{\includegraphics[width=0.4\textwidth]{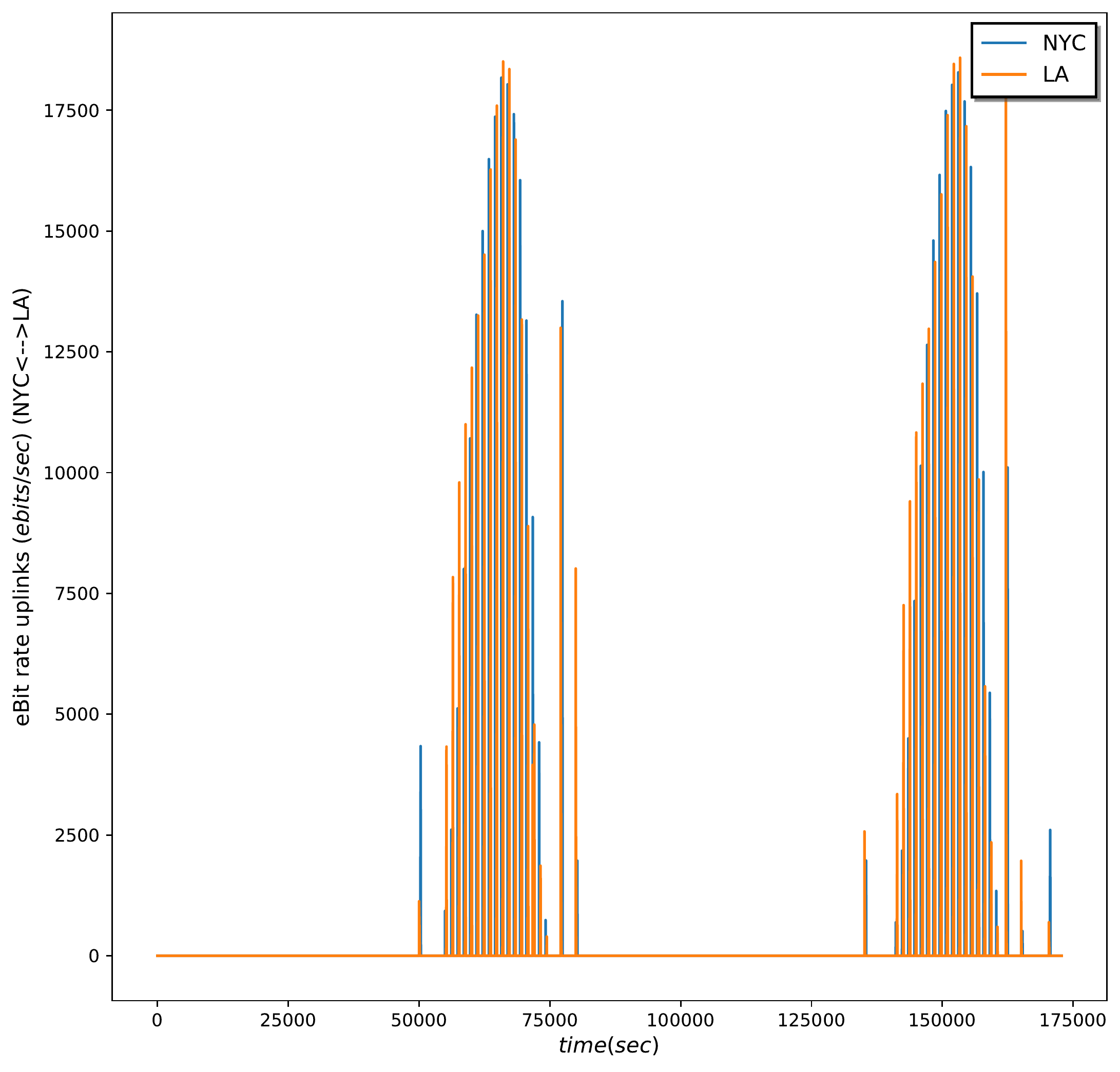}}%
      \label{fig:tau200}%
    }
    \subfloat[Hold over time $\tau = 600$ s.]{%
      \frame{\includegraphics[width=0.4\textwidth]{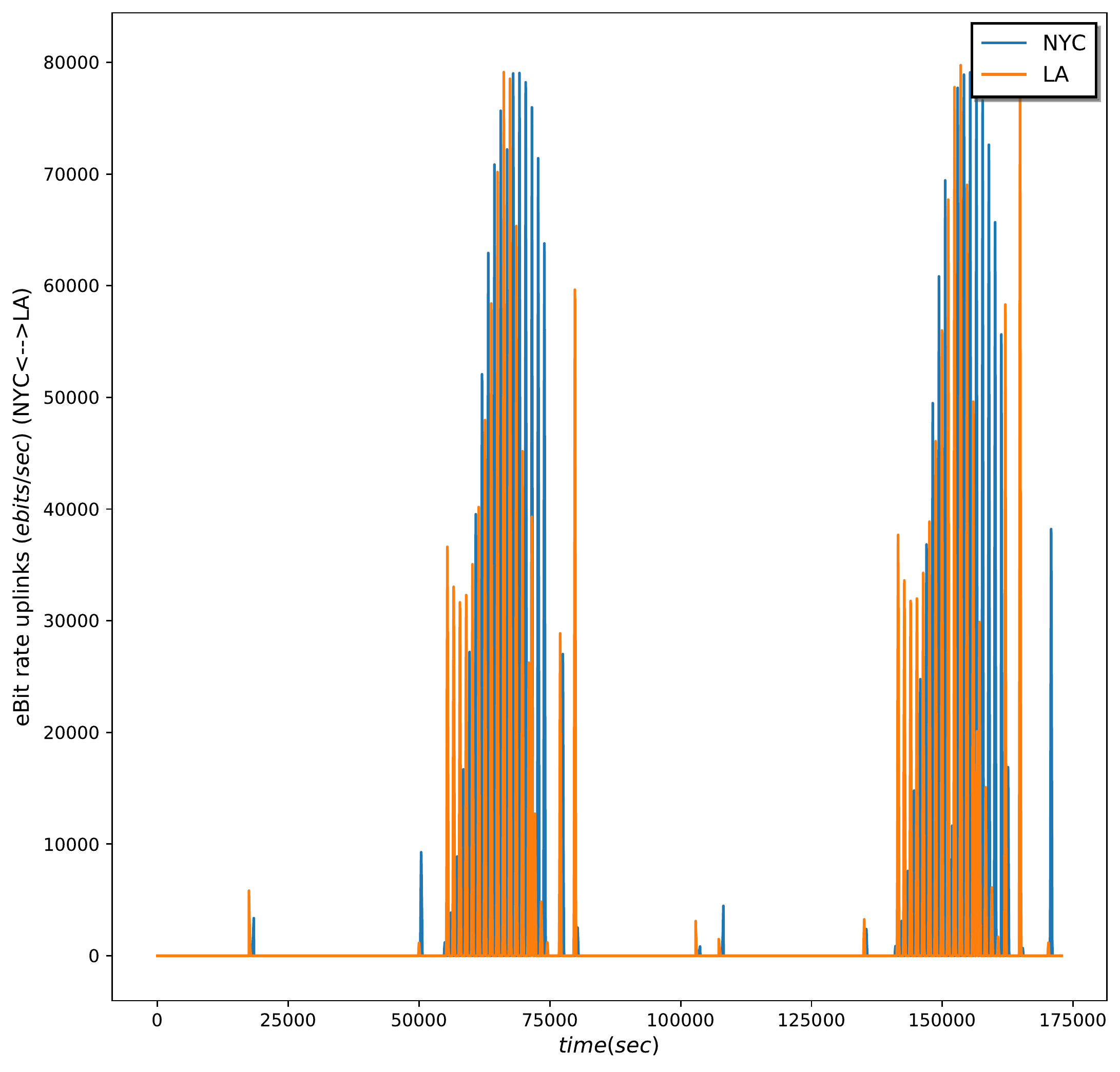}}%
      \label{fig:tau600}%
    }
\caption{Satellite-to-ground sync traces for Los Angles (red line) and New York City (blue line) as the hold over time $\tau$ is varied from $0$ seconds to $600$ seconds. While $\tau$ remains low, it is not possible for these ground stations to synchronize. As clock performance onboard the satellite is improved and $\tau$ increases, the potential for synchronization (ebit rate) as well as the overall time connected increases. (1 orbit of $-50^\circ$ tilt with 5 satellites at 500 km altitude is used for this depiction)}
\label{fig:taueffect}
\end{center}
\end{figure*}

The sync traces for New York City and Los Angeles (for same satellite configuration used in Figure \ref{fig:NYC_LA_uplinks}) is shown in Figure \ref{fig:NYC_LA_sync_uplinks} for a hold over time of 600 seconds (standard Rubidium clocks can hold time at 1 ns precision for around 600 s, even smaller CSACs can do so for around 60-100 s. For reference an ordinary quartz crystal wrist watch can hold time at around 1 millisecond precision for 100 seconds \cite{Lombardi_2008}). Simply put, sync traces are chopped up versions of the connection traces indicating regions of simultaneous connection (or non-simultaneous but within time $\tau$ of each other). Larger $\tau$ implies more common connections (Also see Figure \ref{fig:taueffect}). 

Analysing these connection traces  and sync traces, we can address questions regarding the performance of the time distribution protocol under varying practical constraints. Some of which are: how do the quantum data communication rates change with satellite altitude and other input parameters? If there are resource constraints on the quantity of satellites in orbit and constraints on the quality of their onboard components, to what precision can two ground stations synchronize, and how often can this synchronization be accomplished? Given various satellite configurations (e.g., using polar orbits or different constellation designs), how often do users in specified geographical areas have access to the time distribution service per day? Such questions are complicated to answer generically due to the large parameter space that we need to explore in order to address them and due also to the non-trivial interdependence of these inquiries. In the subsequent sections, we provide specific tools and techniques which we can leverage in order to address these questions and also provide analyses for specific case studies.

\begin{figure}
\begin{center}
\includegraphics[height=8cm]{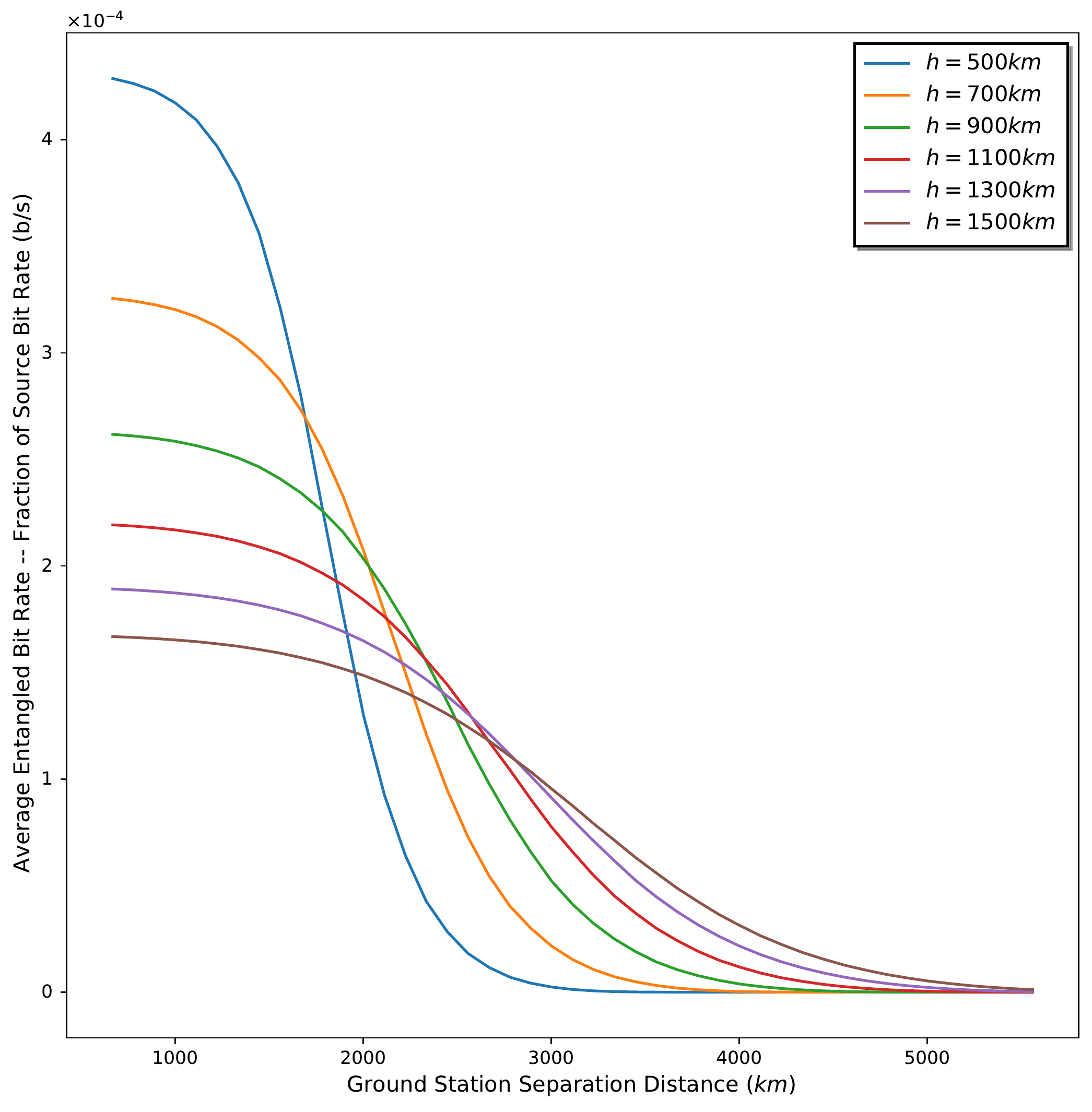}
\caption{Average ebit rate (of the product of two uplink rates) calculated over one day as a function of the arc distance separation between two ground stations for different satellite altitudes, for a value of $\tau = 0$ (poor onboard clock). The two ground stations lie along the equator and the satellite is also in an equatorial orbit. A sharper drop off is seen for low satellite altitudes while for larger altitudes the drop is less pronounced. In this double-link configuration, we see good performance for ground station separation distances at or below 3000 km but a sharp drop in performance above this range.}
\label{fig:ebitdistance}
\end{center}
\end{figure}

We begin our analysis by illustrating the effect of satellite altitude $h$ on the ability to make ground station connections and exchange entangled bits. Network scale is determined by the largest distance between ground stations that can be successfully synced at sub-nanosecond precision. As mentioned earlier, this is achieved whenever communication between the satellite and both ground stations is simultaneously achieved at an ebit rate greater than $\mathcal{R}_c$. Since uplinks are weaker than downlinks, they determine the success of the network. Figure \ref{fig:ebitdistance} summarizes the average entangled bit rate over one day as a function of ground station distance separation for differing satellite altitudes. Two ground stations are placed equatorially some distance $d$ apart with one equatorial satellite placed overhead (as depicted in Figure \ref{fig:duallink}). For a given satellite altitude $h$, the ebit rate averaged over the course of a day is recorded (the product of two uplink ebit rates from each ground station to the satellite is used, which is the same as assuming $\tau = 0$). This procedure iterates with varying ground station separation distances and satellite altitudes to create this figure.
We see for low altitudes more entangled bits are exchanged when the ground stations are relatively close together, as one would expect. But as ground station separation passes $3000$ km, the lowest altitude satellite is no longer able to establish connection and, consequently, no ebits are exchanged (if the average ebit rate is zero, the instantaneous rate must also be zero all through the day). As the orbit is moved to higher altitudes, this does allow for the satellite to connect simultaneously  to more distant ground stations, but at the cost of losing some fraction of ebits due to increased transmission losses. It is clear from Figure \ref{fig:ebitdistance} that there is a trade-off in choosing satellite altitudes that depends on the objective of the timing network. While a lower satellite can effectively deliver a higher bit rate due to smaller transmission losses, the amount of time ground stations can remain in view of the satellite diminishes, particularly as the ground stations become more distant.

\begin{figure}
\begin{center}
\includegraphics[height=8cm]{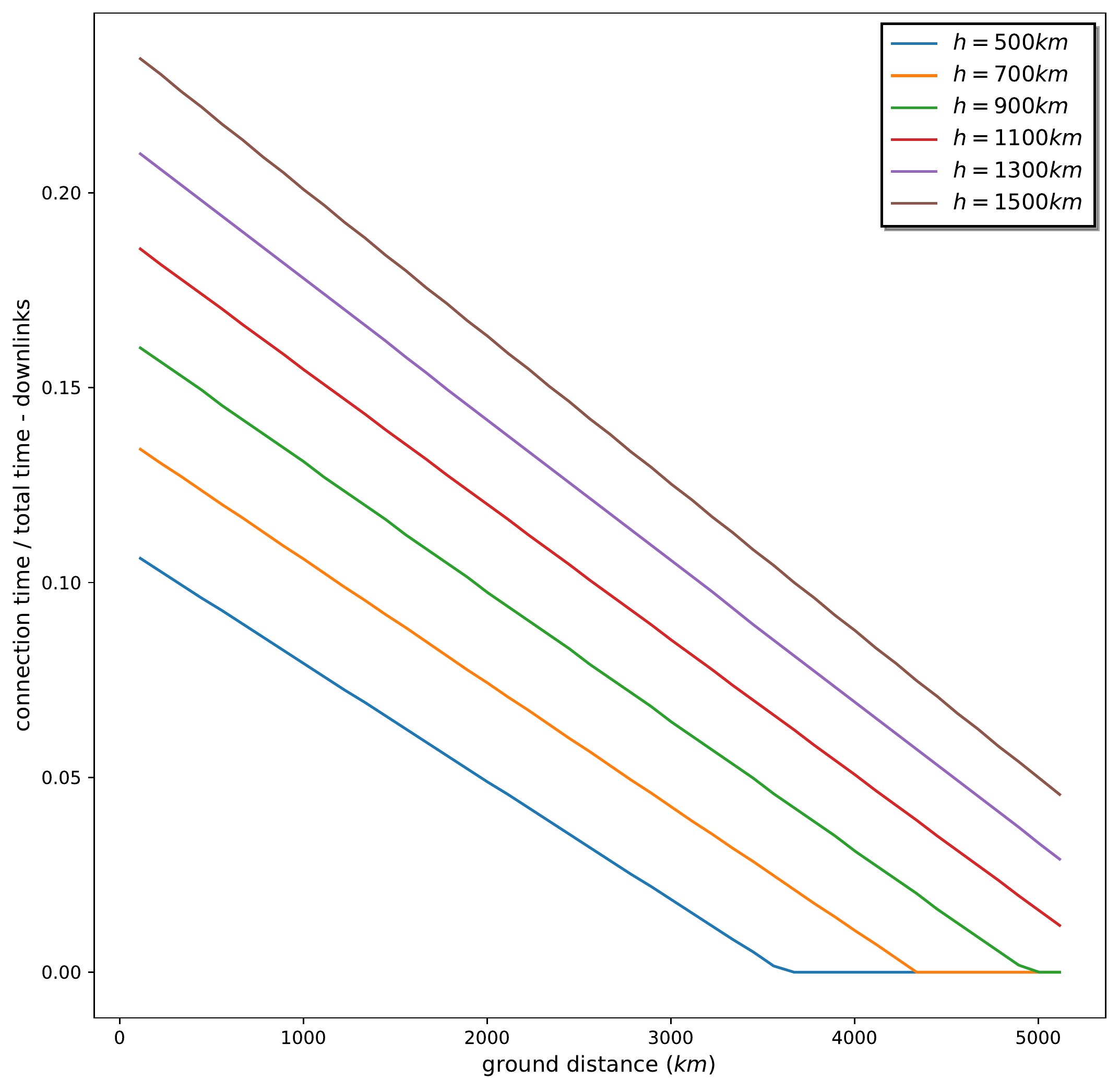}
\caption{Ratio of connected time to the total time for a single orbital pass (${\sim}90$ mins) as a function of ground station arc distance separation for different satellite altitudes. A linear relationship is seen for all altitudes. The average connection time increases with increasing altitude, but average ebit rates fall (Figure \ref{fig:ebitdistance}).}
\label{fig:connectiontime}
\end{center}
\end{figure}

To further expound the effects of satellite constellation altitudes as well as ground station separation distances, we look at Figure \ref{fig:connectiontime} to compare the change in the amount of time the satellite is simultaneously connected to both ground stations
when varying these parameters. As in Figure \ref{fig:ebitdistance}, equatorial ground stations are incrementally separated and attempt synchronisation through an equatorial satellite orbiting at a particular altitude. The ratio of time that the satellite is connected to the two ground stations divided by the total time within one orbital pass (roughly $90$ minutes) is plotted. This is done for several satellite altitudes over a  separation distance between ground stations in the interval $500$ and $5000$ km. Of note here is that the connection time ratio scales linearly as a function of ground station distance up until  connection can no longer be established. Also, as satellite altitude is increased, this time ratio increases as the ground stations continue to stay in view of the satellite for longer.
This highlights that if establishing connections for longer times with at least some minimum number of exchanged entangled bits is favored over having an increased number of exchanged ebits in far apart smaller time windows, satellite constellation altitude is one parameter to consider. 

\begin{figure*}
\begin{center}
    \subfloat[Hold over time $\tau = 0$ s.]{%
      \frame{\includegraphics[height=7cm]{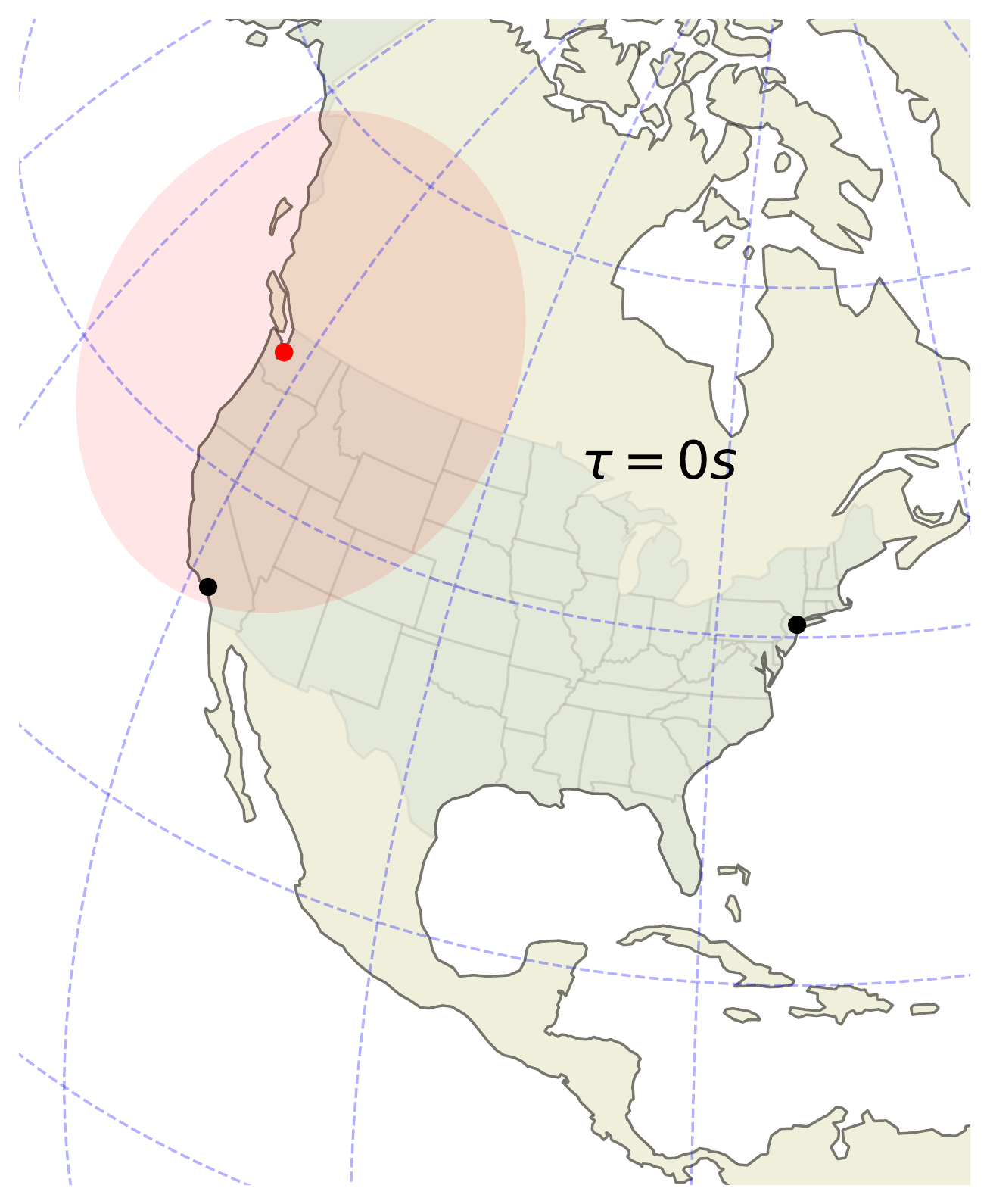}}%
      \label{fig:shadow0}%
    }
    \subfloat[Hold over time $\tau \approx 225$ s.]{%
      \frame{\includegraphics[height=7cm]{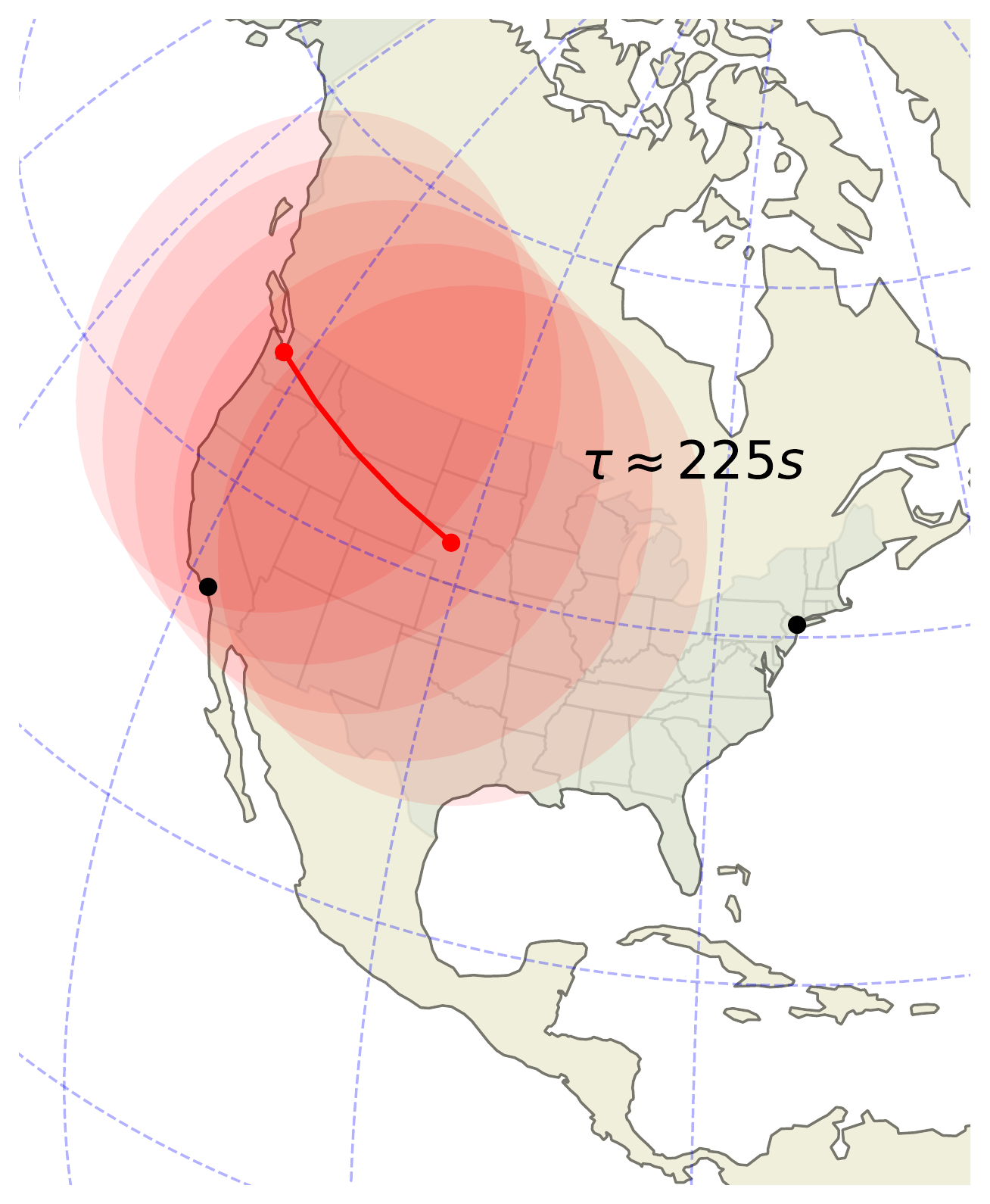}}%
      \label{fig:shadow225}%
    }
    \subfloat[Hold over time $\tau \approx 450$ s.]{%
      \frame{\includegraphics[height=7cm]{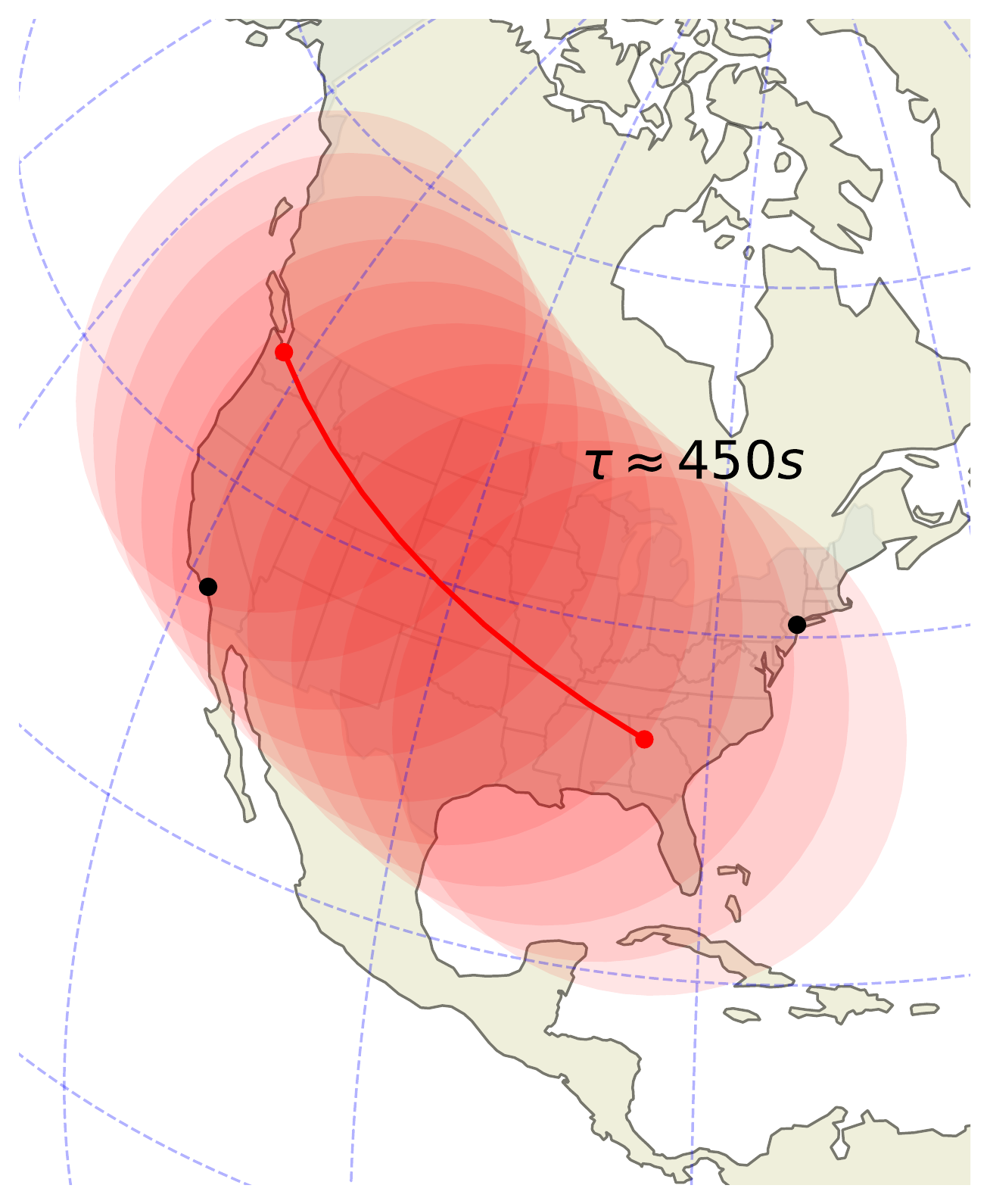}}%
      \label{fig:shadow450}%
    }
\caption{Illustration of the ``shadow'' of a satellite, and the effect of varying hold over time ($\tau$). For the limiting case of no hold over time (\ref{fig:shadow0}), only what the satellite (red dot) has under the instantaneous shadow may be connected and synchronized. As the time window is increased, the satellite track length increases (red line) and includes more distant locations that can then connect and synchronize (\ref{fig:shadow225}). In this representation, as the hold over time is increased to $\tau \approx 450$ s (\ref{fig:shadow450}), ground stations (black dots) LA and NYC can be synchronized with one another. (Figure is not to scale.)}
\label{fig:shadows}
\end{center}
\end{figure*}

It is clear from Figure \ref{fig:ebitdistance} that for a given satellite altitude and cut-off rate there exists a critical distance beyond which two ground stations cannot connect to the same satellite. This motivates us to introduce an intuitive visual tool to understand the previous plots in a simple manner,  alleviating some of the complexity originating from the large number of parameters involved. 
Consider a single satellite. At a given time, it will be able to synchronize with all ground stations for which it has a ebit exchange rate larger than the threshold $\mathcal{R}_c=200$. But since the exchange rate decreases with the satellite-to-ground station distance, this threshold can be translated to a region on the surface of the Earth below the satellite, which we call the ``shadow''. As one moves towards the center of the shadow, concentric regions indicate regions of better sync precision. Hence, higher the cut-off rate $\mathcal{R}_c$ the smaller the shadow. 
With this picture in mind, it is easy to understand that two ground stations would be able to sync at a given level of precision by means of an intermediary satellite whenever they both simultaneously fall under the satellite's shadow. Furthermore, if the clock onboard the satellite is able to hold time at the required precision for an interval $\tau$, the shadow effectively elongates, covering a larger area on the Earth's surface directly proportional to $\tau$. See Figure \ref{fig:shadows}. Since LEO satellites are fast moving ($\approx 8000 m/s$), even small values of $\tau$ translate to large ground coverage areas. For the effect of satellite clock stability on the sync traces see Figs.~\ref{fig:taueffect}.

\begin{figure*}
\begin{center}
    \subfloat[One satellite per orbit.]{%
      \frame{\includegraphics[width=0.4\textwidth]{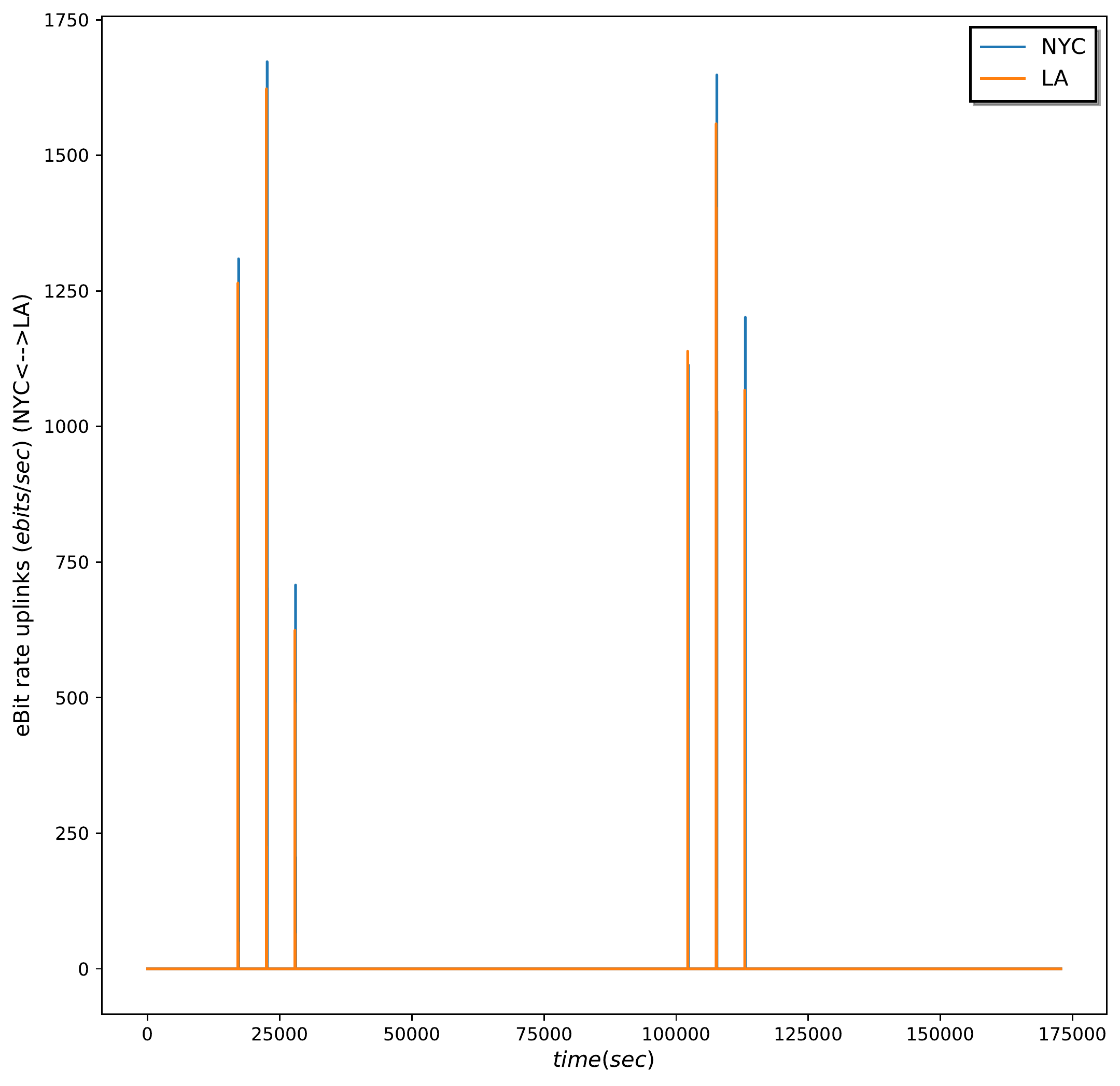}}%
      \label{fig:nsat1}%
    }
    \subfloat[Two satellites per orbit.]{%
      \frame{\includegraphics[width=0.401\textwidth]{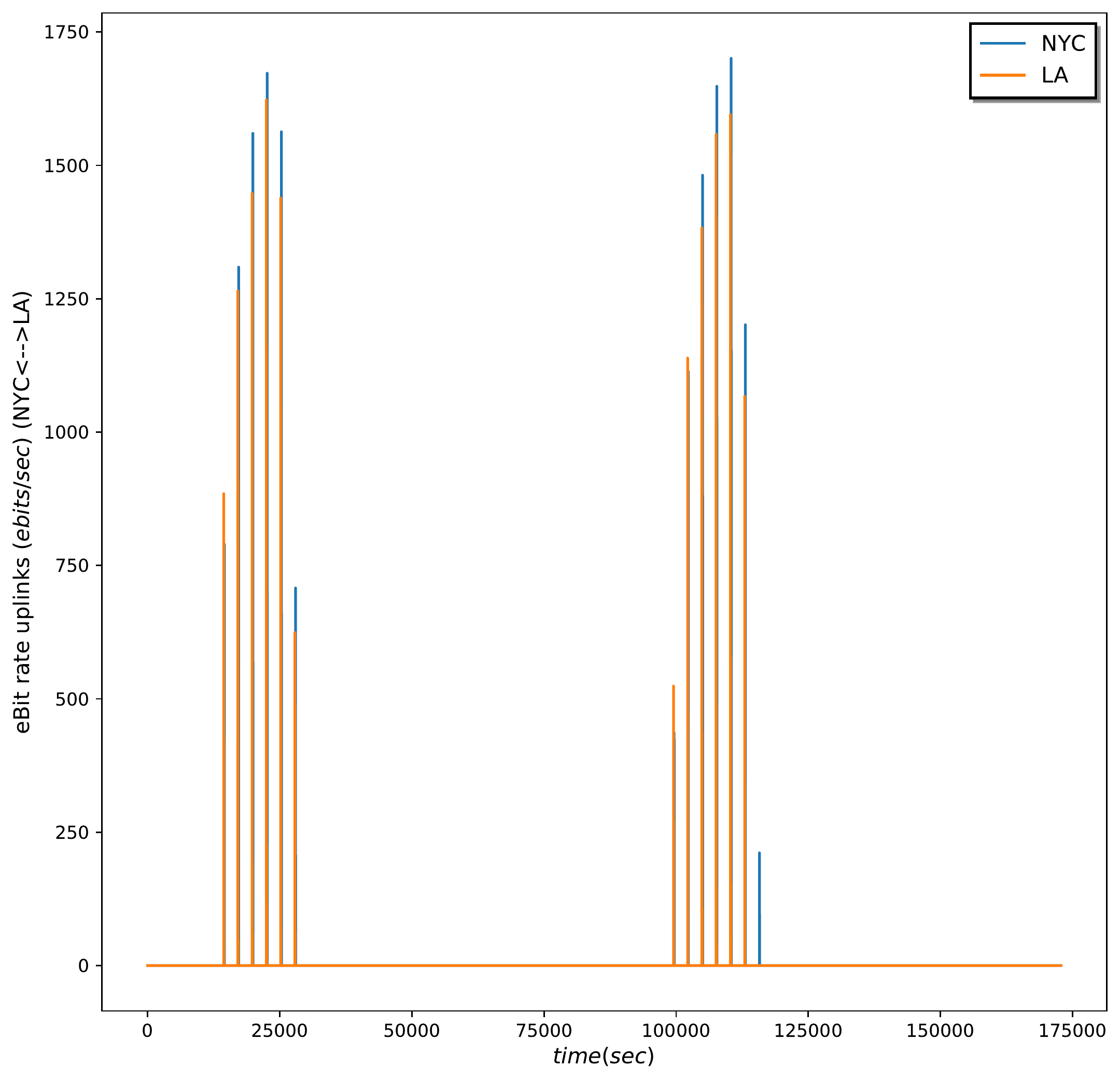}}%
      \label{fig:nsat2}%
    }
    
    \subfloat[Five satellites per orbit.]{%
      \frame{\includegraphics[width=0.4\textwidth]{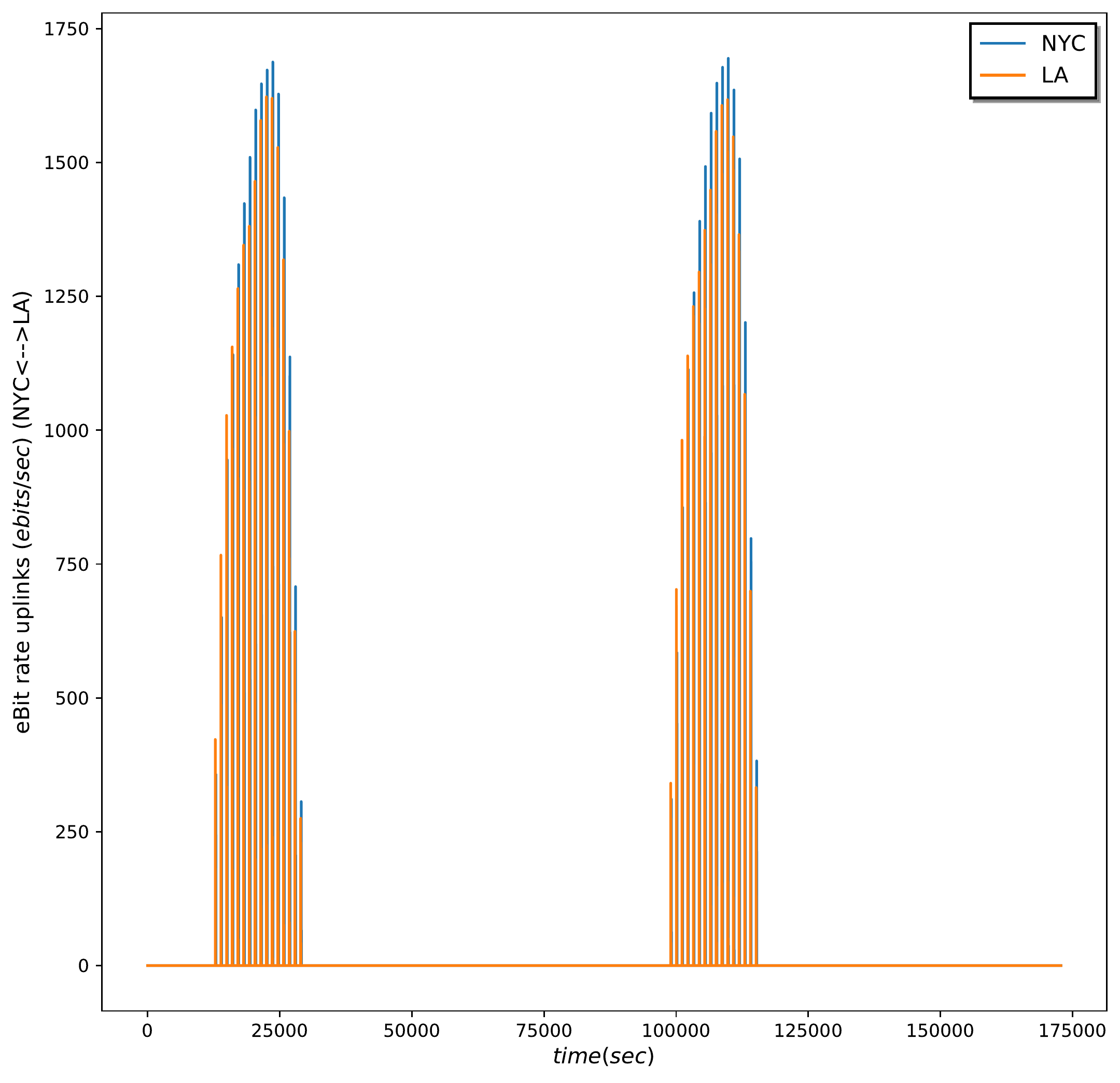}}%
      \label{fig:nsat5}%
    }
    \subfloat[Eight satellites per orbit.]{%
      \frame{\includegraphics[width=0.4\textwidth]{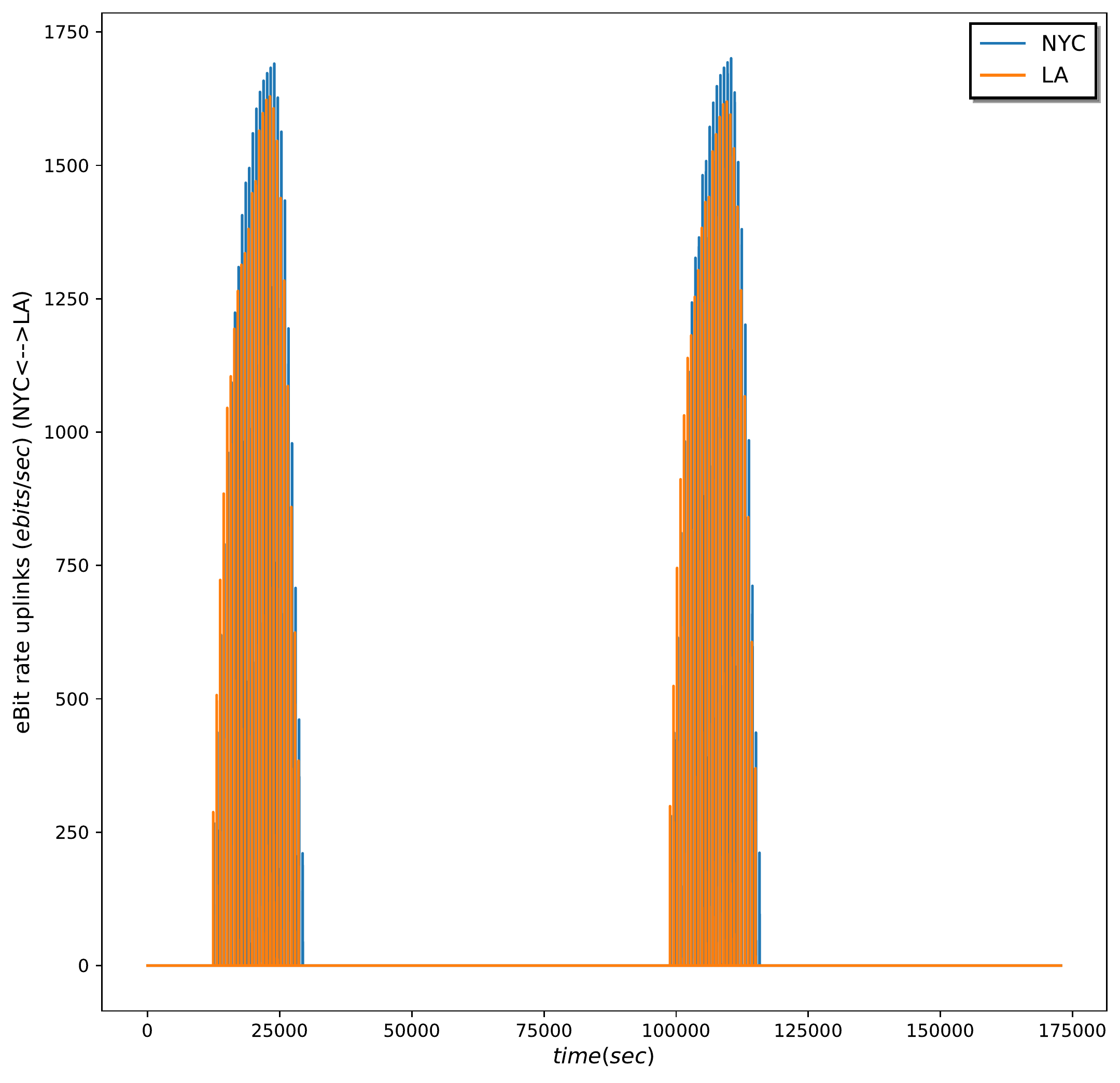}}%
      \label{fig:nsat8}%
    }
\caption{Satellite-to-ground sync traces of New York City and Los Angeles (uplinks) as the number of satellites per orbit (1 orbit of $-50^\circ$ tilt at 500 km altitude in this depiction) is varied from $1$ to $8$. As the number of satellites increase in this particular configuration, the density of viable connections increases; see~\ref{fig:nsat1} to~\ref{fig:nsat2} to~\ref{fig:nsat5}. Beyond a threshold number of satellites (5 in this case), the amount of connection time does not benefit significantly (\ref{fig:nsat8}).}
\label{fig:nsateffect}
\end{center}
\end{figure*}

\begin{figure*}
\begin{center}
    \subfloat[Orbit tilt = $0^\circ$]{%
      \frame{\includegraphics[width=0.4\textwidth]{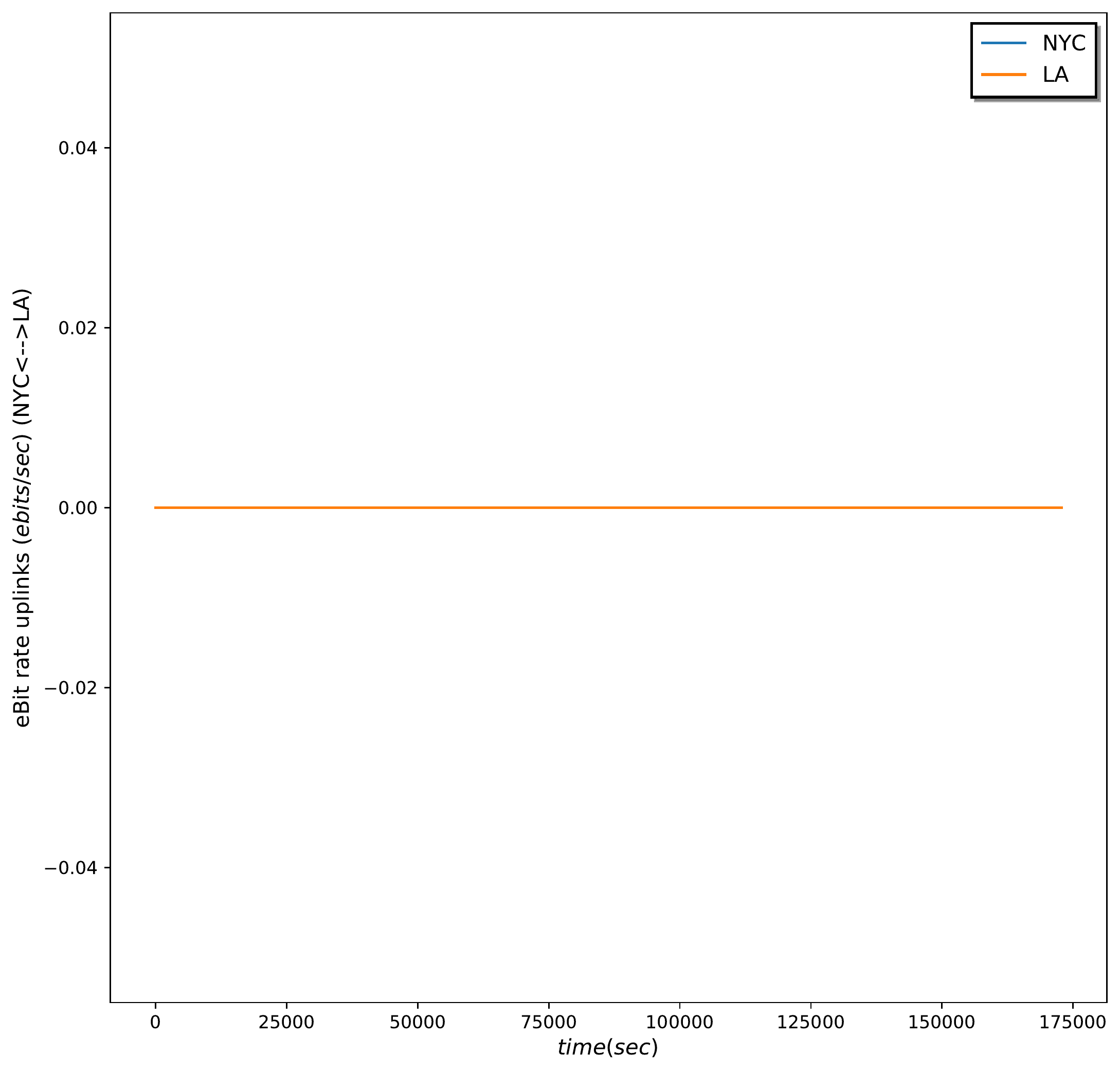}}%
    }
    \subfloat[Orbit tilt = $\pm 25^\circ$]{%
      \frame{\includegraphics[width=0.4\textwidth]{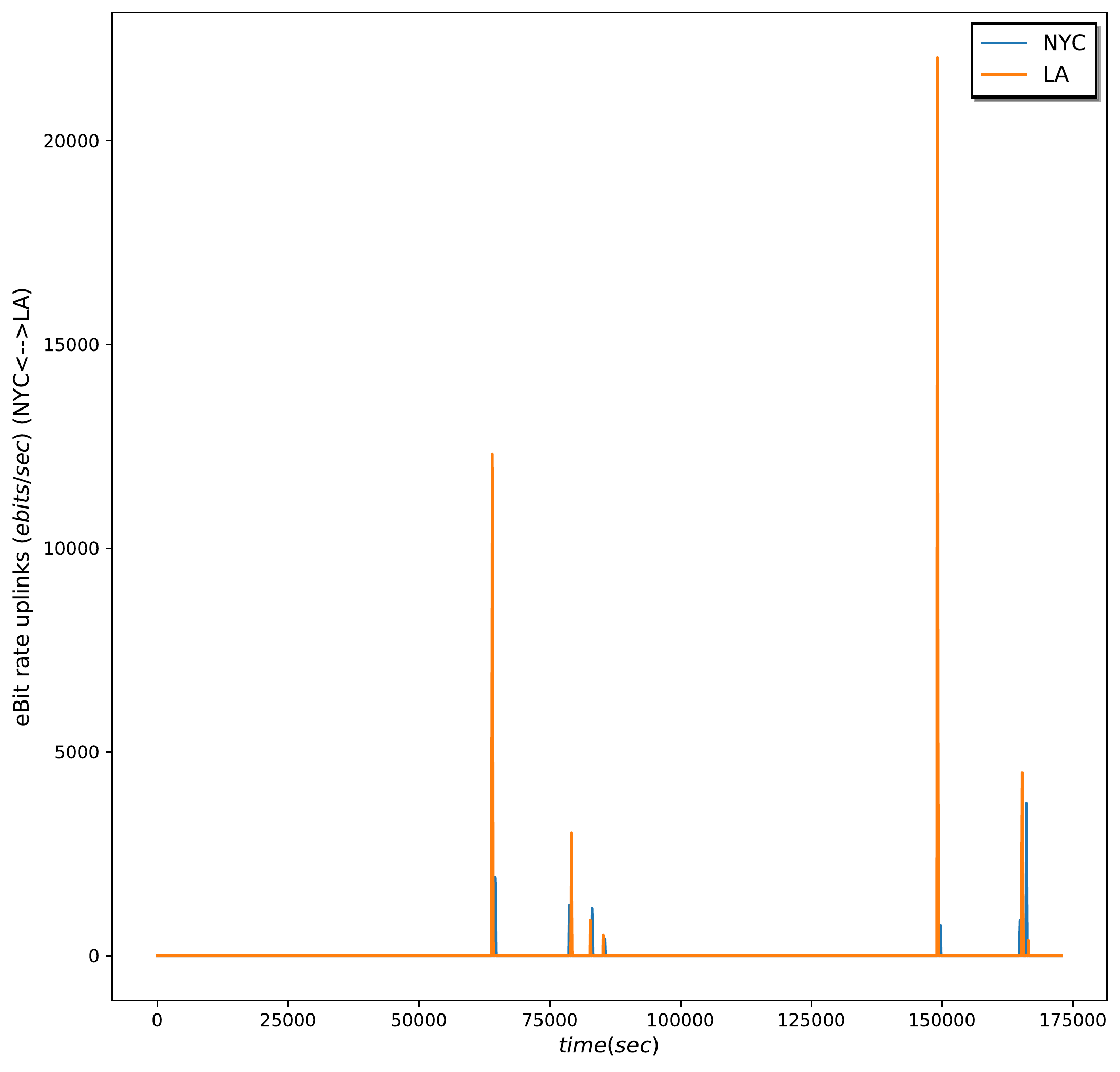}}%
    }
    
    \subfloat[Orbit tilt = $\pm 50^\circ$]{%
      \frame{\includegraphics[width=0.4\textwidth]{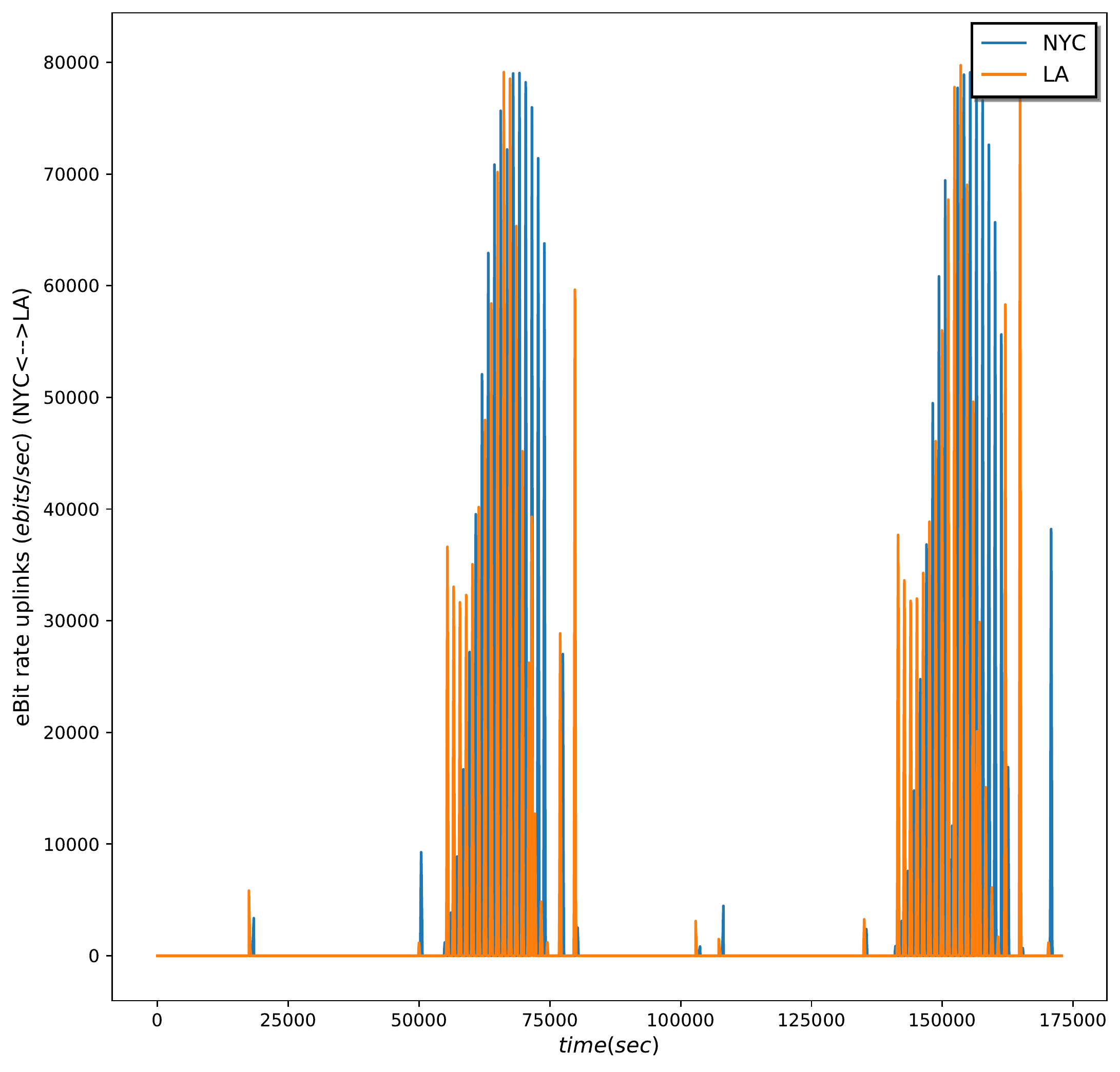}}%
    }
    \subfloat[Orbit tilt = $\pm 75^\circ$]{%
      \frame{\includegraphics[width=0.4\textwidth]{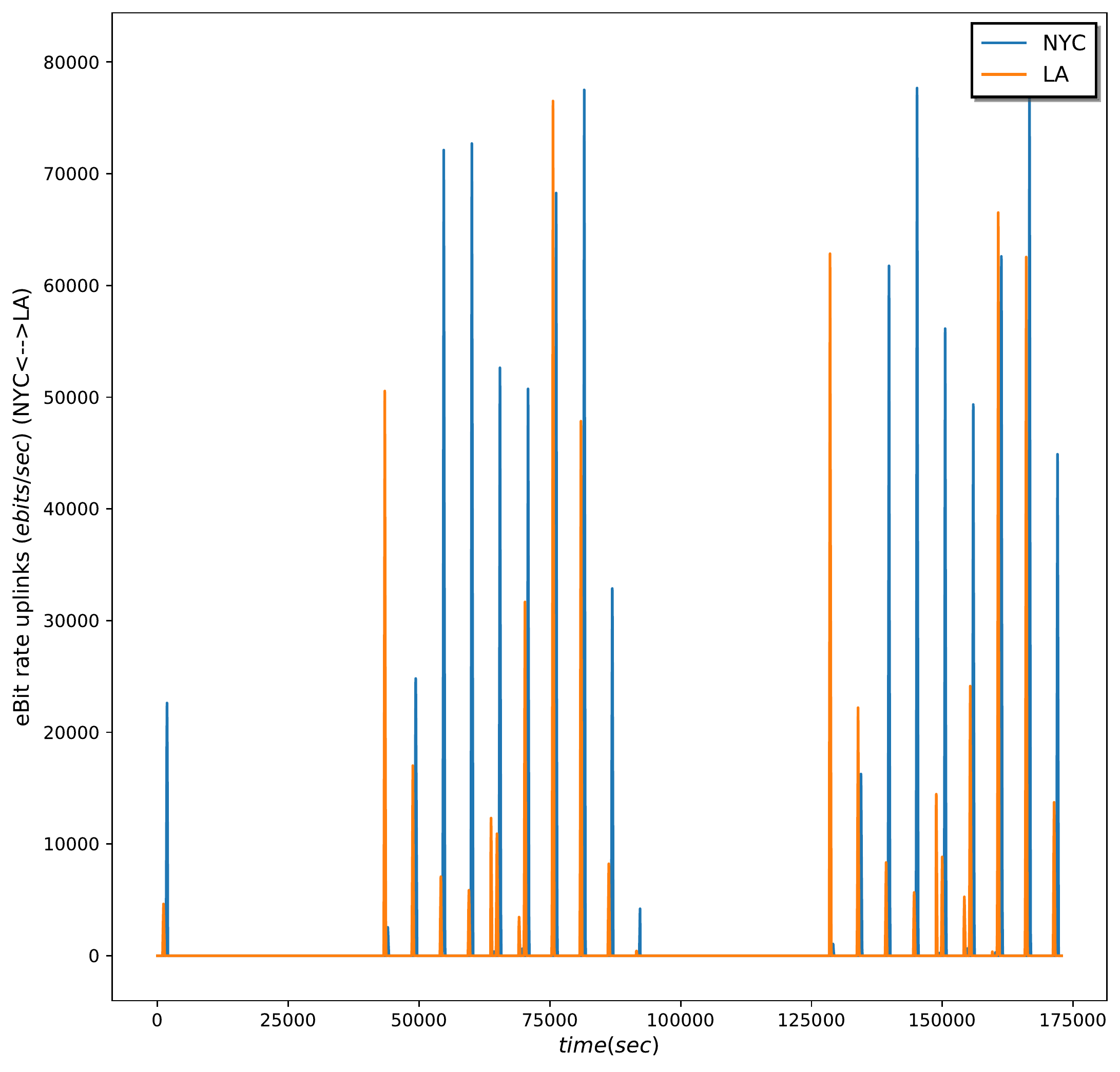}}%
    }
    
\caption{Satellite-to-ground sync traces for New York City-Los Angeles as the orbit tilt is varied from $0^\circ$ to $75^\circ$ (1 orbit with 5 satellites at 500 km altitude). By orienting the satellites orbit along the line joining the two cities, all network figures of merit, including quantum data rates, connection time averages and longest connection gap, are improved. $\pm 50^\circ$ appears to be the optimal tilt for the orbits for this city pair. This is  quantitatively justified below using the concept of the satellite shadow as well.}
\label{fig:tilt_effect}
\end{center}
\end{figure*}

Next up, we briefly discuss the effects of increasing the number of satellites in a particular orbit, the number of orbits utilized, and the orbit type (or inclination from the Earth's axis of rotation) on constellation design. In Figure \ref{fig:nsateffect}, satellite-to-ground sync traces are provided between two ground stations based in Atlanta (ATL) and NYC for a varying number of satellites in a single LEO, over the course of a $2$ day simulation. As the number of satellites is increased, the opportunity for connections to occur also increases up to a threshold. Beyond this point, adding additional satellites for a particular ground station pair does not increase the connection time, although it does allow for higher ebit rates. The particular satellite chosen at each instant to sync the ground stations is based on which provides the least transmission loss at that particular time. Further, in Figure \ref{fig:tilt_effect} we show the effect of orbit inclination on the sync trace between NYC and LA. By orienting the satellite shadow along the line joining the two cities network one can improve outcomes such as quantum data rates, connection time averages, etc. 

The previous discussion illustrates the complexity of the problem of optimization of a network of satellites  arising from the vast space of parameters. The study of the design space are, in part, predicated on which ground stations are important to synchronize, whether some quality of service is to be expected at individual locations, and what type of constraints in terms of costs/resource availability are actually considered in the construction and launch of such a constellation. This leads us, in the next subsection, into considering a concrete network scenario and exploration of time distribution within the imposed constraints.

\subsection{A QCS network for continental US}
\begin{figure*} [ht]
   \begin{center}
   \begin{tabular}{c} 
   \includegraphics[height=8cm]{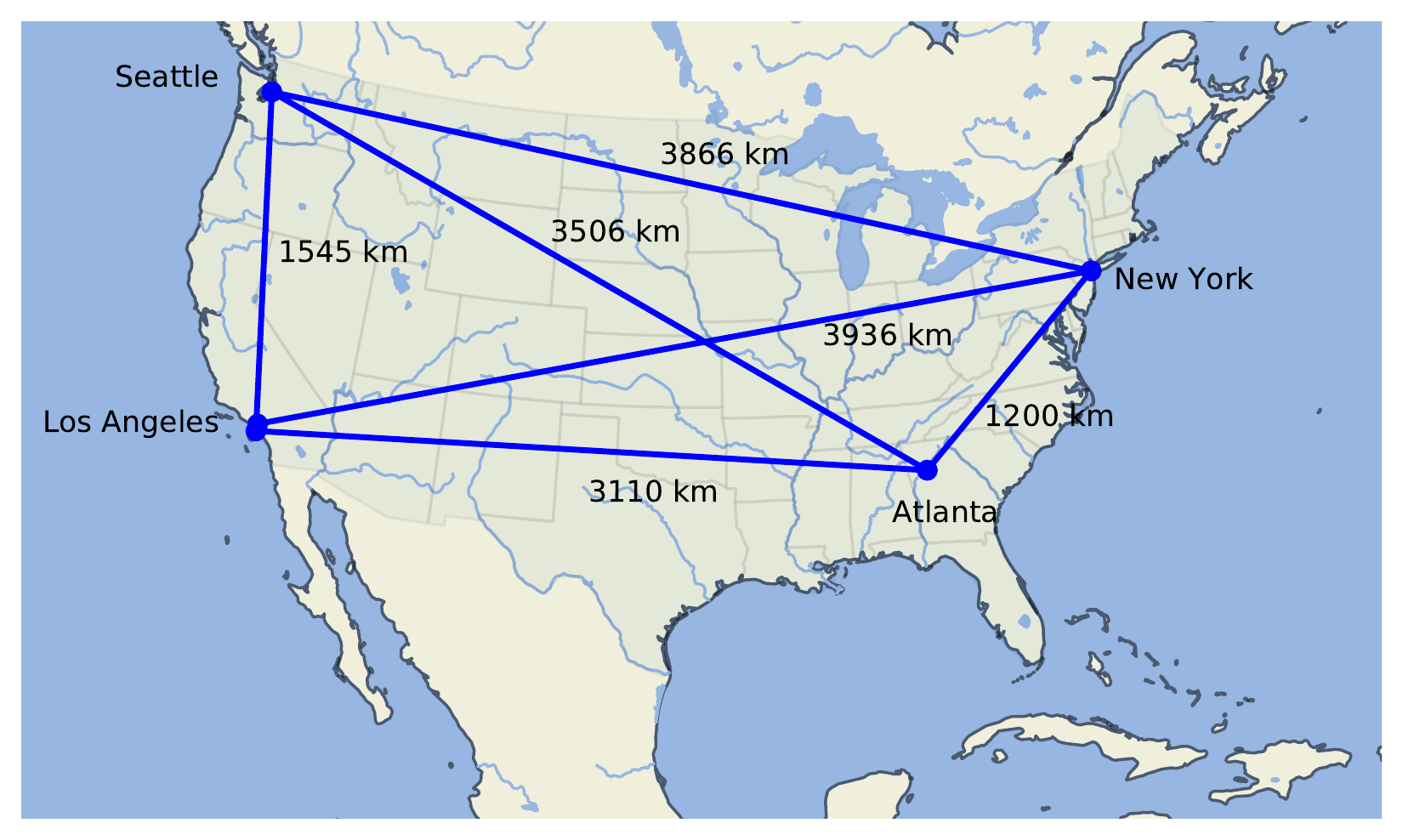}
   \end{tabular}
   \end{center}
   \caption
   {\label{fig:cities}
Selection of representative ground station cities in the contiguous USA. We choose 4 cities of $>1\,\text{M}$ people, separated by close to the maximum possible distance: Seattle, New York, Atlanta and Los Angeles. The inter-city distances (in km) are presented in the map.}
   \end{figure*}

As a concrete scenario, we consider the requirement to synchronize cities lying within the continental United States (see Fig.~\ref{fig:cities}) at a sub-nanosecond precision. The constraints are the availability of moderately stable satellite clocks on-board ($\tau$ of a few minutes), a small number of LEO satellites, and the absence of quantum communication links between satellites (satellites can only communicate to ground stations) 

We assess the performance of our network by the following two criteria:
\begin{itemize}
    \item How well can we synchronize? We  quantify this  by the \emph{average uplink and downlink loss} when a connection is established between a satellite and a ground station. The intuition is straightforward: lower loss leads to a higher number of ebits exchanged, which leads to better statistics for the timing offset calculations; see Section \ref{subsec:extraction}. This all, in turn, leads to higher precision of the QCS protocol. 
    \item How often can we synchronize? Or, similarly, how often do a pair of ground stations get a common satellite in view? The longer the gap between two synchronization events is, the more stable a ground station clock must be to maintain synchronicity. To address this notion of revisits quantitatively, we introduce the \emph{connection time fraction}, calculated as the fraction of time a satellite is connected to the ground station divided by the total simulated time, and the \emph{longest connection time gap}, calculated as the longest time interval a ground station has to stay without a connection to a satellite. The higher the connection time fraction, the higher the total number of ebits shared between the satellite and the ground station. We point out that, when synchronization occurs more frequently (e.g., higher satellite traffic over some terrestrial region), ground stations can leverage lower quality clocks if desired (compared to other ground clocks which sync after long gaps of time and hence must have long stability (hold over) periods). Hence, appropriately addressing questions of revisits can also help to relax certain resource requirements.
\end{itemize}

For this concrete requirement, we assess the feasibility of synchronizing ground stations located at the four corners of contiguous United States (NYC - New York City, LA - Los Angeles, SEA - Seattle, ATL - Atlanta). To begin, let us consider a single satellite in a tilted orbit ($-50^\circ$ angle compared with the Earth's axis of rotation) with a modestly stable onboard clock; $\tau = 500$ s. This constellation configuration, as sparse as it is, does allow for successful quantum clock synchronization among the four cities, as long as the local ground station clocks are sufficiently stable. The average uplink signal loss for ground station pairs range between 27 and 32 decibels. The downlink losses range between 23 and 30 decibels, slightly lower than the uplink losses due to smaller receiving telescope aperture on the satellite \footnote{Also dispersion effects in lower atmosphere lead to beam broadening, this would affect the uplink losses more. We do not consider the effects of turbulence in our simulations.}. For a source emitting $10^7$ entangled pairs per second and a satellite at $500$ km altitude, this translates to between $10,000$ and $15,000$ ebits exchanged per second on average; roughly $10^5$ ebits total per satellite pass. Since we are using a single satellite, the connection time percentages are low and range between $1.3\%$ and $2.8\%$ of a day; i.e., between $20$ and $40$ minutes in total. The longest disconnected intervals between revisits by the satellite range from 15 to 20 hours. So, ground stations whose clocks have a relative frequency drift no bigger than $1 \times 10^{-14}$ connected via this type of constellation ---one satellite--- can provide quantum time transfer capabilities down to $1$ ns precision.
Therefore using our simulations we find that an orbit tilt of around $50^\circ$ about the equatorial plane can provide coverage across the entire contiguous US (also see Figure \ref{fig:tilt_effect}). For larger values of $\tau$, a smaller tilt is required and vice versa. Thus, we show that we are able to extract information about the orbit resources (orbit space and orientation) needed to service specific geographical regions. In effect, this allows us to establish generic network sizes, in terms of geographical area covered by the network. 
This could also be intuitively understood using the shadow picture developed earlier in the section. In this specific case, choosing $\mathcal{R}_c = 200$ ebits/s and $h = 500$ km translates to approximately $35^\circ$ angular diameter for the shadow spot, and the choices of $\tau = 100$s and $\tau = 600$ s give angular lengths of the shadow equal to around $38^\circ$ and $75^\circ$ respectively (which translate approximately to distances of $4000$ km and $8000$ km, respectively). Now, consider the contiguous US: The longitudinal extent of the cities considered here is around $50^\circ$ and the latitudinal extent is only around $15^\circ$. The shadows of polar satellites with no tilts, in this case, will have longitudinal and latitudinal extents of around $35^\circ$ and $75^\circ$ respectively. Hence, an orbit tilt of around $50^\circ$ [$\arccos(50/75)$] about the equatorial plane orients the shadow so as to cover the entire contiguous US.

Next, Table \ref{tab:scenario1results} provides the results for a satellite constellation of $2$ tilted orbits ($50^\circ$ and $-50^\circ$ to the Earth's axis of rotation) of $5$ satellites each with the same ground stations as above. A larger $\tau$ allows not only for lower signal losses but also higher connected time percentages and shorter revisit times. This effect is more pronounced for ground stations with larger separation distances. The takeaway is that the stability of a satellite's onboard clock can markedly enhance the geographical extent to which ground stations can be synchronized. On the other hand, a larger ebit threshold  reduces the connected time percentages and elongates revisit times. 
All these results can be intuitively understood by using the picture of the shadows of each satellite, and the way the shadow changes with both the ebit rate threshold  $\mathcal{R}_c$ and the hold over time $\tau$, already explained in the previous section.

Further, in Table \ref{tab:MEOresults} we summarize the figures of merit for a similar MEO (medium Earth orbit) constellation at 5000 km altitude. Clearly, at the expense of having larger losses, longer connectivity can be obtained. (This, however, should not be thought of as a trade off between sync precision and the size of the coverage area, since  whenever the ebit rate is above the threshold value the required precision can be achieved.) Due to larger instantaneous shadows cast by MEO satellites, in order to cover a similar geographical area (contiguous US in this case) the requirement on $\tau$ is much smaller (lower stability satellite clocks can be used).

\begin{table*}
\begin{center}
\begin{tabular}{|p{2.2cm}|p{2.3cm}|p{2.4cm}|p{2.5cm}|p{2.2cm}|}
\hline
\multicolumn{5}{|c|}{$h = 500$ km, $\mathcal{R}_c = 200$ ebits/s, $\tau = 100$ s} \\
\hline
Ground Station Pair (GS1/GS2) & Avg. uplink ebit loss(dB/dB) & Avg. downlink ebit loss (dB/dB) & Percent of day connected & Longest connection gap (hours)  \\
\hline
NYC/LA  & 42/42 & 41/41 & 0.6\%   & 17 \\
NYC/SEA & 44/44 & 43/43 & 0.1\% & 15 \\
NYC/ATL & 27/28 & 23/24 & 19\%  & 6  \\
LA/SEA  & 28/29 & 24/25 & 16\%  & 6  \\
LA/ATL  & 33/33 & 31/31 & 5\%   & 6  \\
SEA/ATL & 39/37 & 37/36 & 1.5\%   & 11  \\
\hline
\multicolumn{5}{|c|}{$\mathcal{R}_c = 200$ ebits/s, $\tau = 200$ s} \\
\hline
NYC/LA  & 33/33 & 23/24 & 3\% & 10 \\
NYC/SEA & 36/38 & 24/30 & 2\% & 16 \\
NYC/ATL & 27/28 & 24/24 & 21\% & 6  \\
LA/SEA  & 28/29 & 24/26 & 19\% & 6  \\
LA/ATL  & 28/28 & 24/25 & 8\% & 6  \\
SEA/ATL & 33/30 & 29/25 & 4\% & 8  \\
\hline
\multicolumn{5}{|c|}{$\mathcal{R}_c = 200$ ebits/s, $\tau = 500$ s} \\
\hline
NYC/LA  & 27/28 & 23/24 & 11\% & 16 \\
NYC/SEA & 28/31 & 23/29 & 8\% & 16 \\
NYC/ATL & 27/28 & 23/24 & 21\% & 6  \\
LA/SEA  & 28/29 & 24/25 & 19\% & 6  \\
LA/ATL  & 28/28 & 24/24 & 15\% & 6  \\
SEA/ATL & 31/28 & 27/24 & 12\% & 10  \\
\hline

\multicolumn{5}{|c|}{$\mathcal{R}_c = 500$ ebits/s, $\tau = 100$ s} \\
\hline
NYC/LA  & 42/42 & 41/41 & 0.1\%   & 21 \\
NYC/SEA & $\infty$/$\infty$ & $\infty$/$\infty$ & 0\% & $\infty$ \\
NYC/ATL & 26/27 & 22/23 & 17\%  & 6  \\
LA/SEA  & 28/28 & 24/25 & 15\%  & 6  \\
LA/ATL  & 33/33 & 31/31 & 3\%   & 7  \\
SEA/ATL & 38/37 & 37/36 & 0.8\%   & 18  \\
\hline
\multicolumn{5}{|c|}{$\mathcal{R}_c = 500$ ebits/s, $\tau = 200$ s} \\
\hline
NYC/LA  & 33/33 & 31/31 & 2\% & 17 \\
NYC/SEA & 36/37 & 35/36 & 1\% & 16 \\
NYC/ATL & 27/28 & 23/24 & 19\% & 6  \\
LA/SEA  & 28/28 & 24/25 & 17\% & 6  \\
LA/ATL  & 28/28 & 25/25 & 7\% & 7  \\
SEA/ATL & 33/30 & 31/28 & 3\% & 17  \\
\hline
\multicolumn{5}{|c|}{$\mathcal{R}_c = 500$ ebits/s, $\tau = 500$ s} \\
\hline
NYC/LA  & 27/27 & 23/24 & 9\% & 17 \\
NYC/SEA & 27/31 & 23/29 & 7\% & 16 \\
NYC/ATL & 27/28 & 23/24 & 19\% & 6  \\
LA/SEA  & 28/28 & 24/25 & 17\% & 6  \\
LA/ATL  & 27/28 & 24/24 & 13\% & 7 \\
SEA/ATL & 30/28 & 27/24 & 10\% & 10  \\
\hline
\end{tabular}

\caption{QCS network figures of merit for ground station pairs in the contiguous US with a constellation of 10 LEO satellites (2 tilted polar orbits of 5 satellites each). Orbit altitudes are set to $500$ km, ebit connection rate threshold $\mathcal{R}_c$ and hold over time $\tau$ are varied. Average ebit losses for up and downlink are reported from each ground station in decibels (this includes detector inefficiencies $\approx 6$ dB). An average loss of 30 decibels converts to a ebit loss fraction of $10^{-3}$. Therefore, if the entangled source rate is $10^7$ ebits/s then, on average, $10,000$ ebits are detected between the satellite and ground station per second when a connection is established.}
\label{tab:scenario1results}
\end{center}
\end{table*}

\begin{table*}
\begin{center}
\begin{tabular}{|p{2.2cm}|p{2.3cm}|p{2.4cm}|p{2.5cm}|p{2.2cm}|}
\hline
\multicolumn{5}{|c|}{$h = 5000$ km, $\mathcal{R}_c = 200$ ebits/s, $\tau = 100$ s} \\
\hline
Ground Station Pair (GS1/GS2) & Avg. uplink ebit loss(dB/dB) & Avg. downlink ebit loss (dB/dB) & Percent of day connected & Longest connection gap (hours)  \\
\hline
NYC/LA  & 37/38 & 37/37 & 60\%   & 4 \\
NYC/SEA & 37/38 & 37/38 & 60\% & 3 \\
NYC/ATL & 37/37 & 37/37 & 89\%  & 1  \\
LA/SEA  & 38/38 & 38/37 & 91\%  & 0.8  \\
LA/ATL  & 37/37 & 37/37 & 68\%   & 3  \\
SEA/ATL & 38/37 & 37/36 & 66\%   & 3  \\
\hline
\end{tabular}

\caption{QCS network figures of merit for ground station pairs in the contiguous US with a constellation of 10 MEO satellites (2 tilted polar orbits of 5 satellites each). Orbit altitudes are set to $5000$ km, ebit connection rate threshold $\mathcal{R}_c$ at 200 ebits/s and hold over time $\tau$ at 100s. Average ebit losses for uplink and downlink are reported from each ground station in decibels (this includes detector inefficiencies $\approx 6$ dB). An average loss of 30 decibels converts to a ebit loss fraction of $10^{-3}$. Therefore, if the entangled source rate is $10^7$ ebits/s then, on average, $10,000$ ebits are detected between the satellite and ground station per second when a connection is established. Figures of merit show considerable increase in the connectivity even for a modest hold over time. Also the losses are more uniform, all cities now operate at around the same precision as against the LEO case in Table \ref{tab:scenario1results} where the losses range from 26-44 dB. Although some cities now receive lower ebit rates than the LEO case, setting the cut-off rate guarantees better than 1 nanosecond precision.}
\label{tab:MEOresults}
\end{center}
\end{table*}

\subsection{Static simulation} \label{subsec:static_simulations}

The goal of these static simulations will be to quantitatively determine   a practical lower bound ({\em ebit threshold}) ${\cal{R}}_c$  for the ebit rate in order to yield a clock offset estimate at a certain precision. The simulation is static in the sense that the relative motion between the satellite and ground station is not included; rather, an appropriate fixed value for channel transmissivity is chosen (acquisition time required to perform the sync protocol is small ($T_a \approx 250 ms$), the link distance and hence the losses do not change at this time scale). Monte Carlo simulations are performed to generate photon time stamps and thereby the correlation functions defined in section \ref{subsec:extraction}. However, relative clock drift can be included in the simulation.

The static simulation of the performance of the QCS protocol uses a Monte Carlo simulation of the photon pair production, the link loss, photon detection, and detection timestamping. The simulation parameters include the time correlated photon pair source rate (Poisson distributed pair production with a specific rate), the (fixed) losses in the optical links (dB), detector efficiencies (fixed, no dead-time is modelled), detector dark count rates (Hz), timing jitter (ps, FWHM), timestamp resolution (ps), the clock drift rate for each clock (fractional frequency accuracy), and the total acquisition time during which the photons are detected (seconds). For each set of these parameters, a different true clock offset is uniformly randomly chosen in the interval between 0 and 1 microsecond and 100 different timestamp datasets are generated for each value. Thus, we simulate 100 instances each for a range of QCS scenarios each with a fixed clock offset. Once the time stamps are generated and detected in each instance, fast Fourier transforms (FFTs) of the resulting times series are performed to find the required cross correlation functions. `Successful' cases are those whose simulated clock offset estimate is within 1 ns of the true value.

Although the simulation is static in that there is no relative motion between the two clocks, a constant velocity between them will have the effect of inducing an apparent relative clock drift between the clocks. The relative velocity between ground stations and satellites can be considered constant if the required acquisition time is small enough. Therefore, if this relative velocity is known to sufficient precision, the effect of motion can be corrected by applying a compensating transformation to the recorded timestamp data. Alternatively, with a sufficient data rate one can estimate both the clock offset and relative clock drift rate directly from the timestamp data. In this way, only limited external information about the satellite's motion would need to be used. This would be preferred in the case that the time distribution network functions independently of GPS or other GNSS constellation. For more details about these techniques we refer the reader to \cite{ho:09}.

Note that the acquisition time ($T_a$) used for each independent clock offset estimate is limited by the effective (uncompensated motion induced plus intrinsic) relative clock drift rate. This is because if the effective drift rate times the acquisition time exceeds the timestamp resolution, the height of the cross-correlation peaks rapidly begin to be reduced since the correlations become spread over many time bins in the FFT window used for the cross-correlation calculations. In the simulations we have chosen an upper limit of $T_a=250$ milliseconds. For a timestamp resolution of 50 picoseconds, this means the clock synchronization algorithm will have no difficulty with a relative fractional frequency offset of $2 \times 10^{-10}$ or less, as determined by the fractional frequency error of the clock. For reference, the frequency stability of a CSAC is about $3 \times 10^{-10}$ measured over 1 second \cite{Microchip}. Therefore, assuming accurate enough compensation for the relative motion, the simulations results below should be consistent with those achievable when using clocks with short-term stability no better than provided by CSACs. Note that the frequency stability is measured in the time domain by the Allen Deviation (ADEV) which is the uncertainty of the clock's frequency measured over a specified length of time. Since the acquisition time used in the simulations are on the order of 1 second or less, the ADEV at 1 second provides a useful upper bound on the frequency error as measured by the reference clock. 

The results of the simulations with fixed acquisition time of 250 ms and varying link losses are given in Tables \ref{tab:static_effect_no_jitter_50ps_res} -- \ref{tab:static_effect_loss_100ps_jitter_150ps_res}. The parameters have been chosen to be representative of realistic components. The simulation sets each of the detector's efficiency to 50\% with a dark count rate of 1000 Hz. We see in Table \ref{tab:static_effect_no_jitter_50ps_res} that in the limiting case of no timing jitter (due to detectors and timestamper) and a timestamper resolution of 50 ps, there is a sharp increase in failure rate of the clock offset estimation as the ebit rate drops below about 100 ebits/s. As can be seen in Table \ref{tab:static_effect_loss_100ps_jitter_50ps_res}, with a realistic value of 100 ps for the timing jitter, the required ebit rate increases to about 200 ebits/s. When the number of exchanged ebits falls much below 50, the clock synchronization algorithm begins to fail at least as often as it succeeds. Thus, for the chosen realistic parameters, an appropriate cut-off for the ebit rate is about 200 ebits/s with an acquisition time of 250 to 500 milliseconds. In Table \ref{tab:static_effect_acquisition} we see the effect of changing the acquisition time for each offset estimate for a fixed value of the link loss. As the acquisition time is reduced from 250 ms, the number of successful offset estimates observed over the 100 simulations is reduced, while for longer acquisition times such as 500 ms, the effect of the frequency instability of the satellite clock increases the width of the cross correlation peak resulting in a SNR that is roughly identical to the 250 ms case and with reduced accuracy. 

In order to increase the achievable clock synchronization accuracy, the individual estimates from each acquisition window can be averaged. This will reduce the random errors due to timing jitter and other random system noise as a function of the integration time. While for LEO satellites this integration time will be limited to a few hundred seconds, for many locations this should allow the synchronization accuracy to approach the timestamp resolution. The stability of the clock on the satellite will also determine the length of optimal integration time since the uncertainty in the clock's drift rate will place an upper bound on the optimum integration time.

\begin{table*}[ht]
\begin{center}
\begin{tabular}{ |p{2cm}|p{1.5cm}|p{1.5cm}|p{2.0cm}|p{2.7cm}|p{3.0cm}| } 
\hline
Link Photon Loss (dB) & Success Rate (\%) & Mean ebit Rate (ebits/s) & Mean Cross Correlation SNR & Mean Clock Offset Error (ps) & Mean Offset Error `Successful' cases (ps) \\
\hline
34.0 & 100 & 995 & 76.8 $\pm$ 6.0 & 29 $\pm$ 0 & Same \\ 
36.0 & 100 & 628 & 59.2 $\pm$ 5.4 & 28 $\pm$ 0 & Same \\ 
38.0 & 100 & 396 & 44.7 $\pm$ 5.2 & 39 $\pm$ 0 & Same \\ 
40.0 & 100 & 250 & 28.2 $\pm$ 4.3 & 48 $\pm$ 16 & Same \\
42.0 & 100 & 158 & 19.6 $\pm$ 3.3 & 36 $\pm$ 25 & Same \\
44.0 & 97 & 100 & 15.2 $\pm$ 3.4 & (0.2 $\pm$ 1.2) $\times 10^6$ & 43 $\pm$ 0 \\
46.0 & 54 & 63 & 11.1 $\pm$ 2.1 & (2.3 $\pm$ 7.8) $\times 10^6$ & 33 $\pm$ 0 \\
\hline
\end{tabular}
\caption{Static simulation results: Varying link loss with fixed acquisition time ($T_a = 250$ milliseconds), 50 ps timestamp resolution, no timing jitter, satellite clock frequency accuracy of $3 \times 10^{-10}$, and 1000 Hz dark count rate with 50\% detection efficiency per detector. The simulation includes no loss between the photon pair source and local detectors. `Successful' cases are those whose simulated clock offset estimate is within 1 ns of the true value.}
\label{tab:static_effect_no_jitter_50ps_res}
\end{center}
\end{table*}

\begin{table*}[ht]
\begin{center}
\begin{tabular}{ |p{2cm}|p{1.5cm}|p{1.5cm}|p{2.0cm}|p{2.7cm}|p{3.0cm}| } 
\hline
Link Photon Loss (dB) & Success Rate (\%) & Mean ebit Rate (ebits/s) & Mean Cross Correlation SNR & Mean Clock Offset Error (ps) & Mean Offset Error `Successful' cases (ps) \\
\hline
34.0 & 100 & 995 & 23.2 $\pm$ 2.5 & 42 $\pm$ 42 & Same \\ 
36.0 & 100 & 628 & 18.8 $\pm$ 2.7 & 42 $\pm$ 45 & Same \\
38.0 & 100 & 396 & 14.5 $\pm$ 2.3 & 43 $\pm$ 53 & Same  \\
40.0 & 80 & 250 & 11.3 $\pm$ 1.9 & $(0.7 \pm 5.3) \times 10^6$ & 47 $\pm$ 51 \\
41.0 & 67 & 199 & 10.4 $\pm$ 1.6 & $(0.5 \pm 10.1) \times 10^6$ & 25 $\pm$ 60 \\
42.0 & 35 & 158 & 9.8 $\pm$ 1.3 & $(1.4 \pm 9.2) \times 10^6$ & 22 $\pm$ 39 \\
44.0 & 1 & 100 & 9.2 $\pm$ 0.7 & $(3.7 \pm 13.6) \times 10^6$ & 145 \\
\hline
\end{tabular}
\caption{Static simulation results: Varying link loss with fixed acquisition time ($T_a = 250$ milliseconds), 50 ps timestamp resolution, 100 ps detector timing jitter (FWHM), satellite clock frequency accuracy of $3 \times 10^{-10}$, and 1000 Hz dark count rate with 50\% detection efficiency per detector. The simulation includes no loss between the photon pair source and local detectors. `Successful' cases are those whose simulated clock offset estimate is within 1 ns of the true value.}
\label{tab:static_effect_loss_100ps_jitter_50ps_res}
\end{center}
\end{table*}

\begin{table*}[ht]
\begin{center}
\begin{tabular}{ |p{1.7cm}|p{1.5cm}|p{1.5cm}|p{2.0cm}|p{2.7cm}|p{3.0cm}| } 
\hline
Link Photon Loss (dB) & Success Rate (\%) & Mean ebit Rate (ebits/s) & Mean Cross Correlation SNR & Mean Clock Offset Error (ps) & Mean Offset Error `Successful' cases (ps)\\
\hline
34.0 & 100 & 995 & 14.9 $\pm$ 1.6 & 39 $\pm$ 82 & Same \\ 
36.0 & 100 & 628 & 12.2 $\pm$ 1.7 & 59 $\pm$ 105 & Same \\
38.0 & 98 & 396 & 9.8 $\pm$ 1.3 & (0.5 $\pm$ 3.9)$\times 10^6$ & 43 $\pm$ 97 \\
40.0 & 54 & 250 & 8.3 $\pm$ 1.1 & (0.3 $\pm$ 1.2)$\times 10^7$ & 52 $\pm$ 122 \\
41.0 & 26 & 199 & 7.9 $\pm$ 0.8 & (0.7 $\pm$ 18.2)$\times 10^6$ & 47 $\pm$ 158 \\
42.0 & 10 & 158 & 7.9 $\pm$ 0.6 & (0.5 $\pm$ 2.2)$\times 10^7$ & 14 $\pm$ 118 \\
44.0 & 2 & 100 & 7.9 $\pm$ 0.5 & (0.5 $\pm$ 2.0)$\times 10^7$ & 128 $\pm$ 0 \\
\hline
\end{tabular}
\caption{Static simulation results: Varying link loss with fixed acquisition time ($T_a = 250$ milliseconds), 100 ps timestamp resolution, 200 ps detector timing jitter (FWHM), satellite clock frequency accuracy of $3 \times 10^{-10}$, and 1000 Hz dark count rate with 50\% detection efficiency per detector. The simulation includes no loss between the photon pair source and local detectors. `Successful' cases are those whose simulated clock offset estimate is within 1 ns of the true value.}
\label{tab:static_effect_loss_100ps_jitter_150ps_res}
\end{center}
\end{table*}

\begin{table*}[ht]
\begin{center}
\begin{tabular}{ |p{2cm}|p{1.5cm}|p{1.5cm}|p{1.5cm}|p{2.0cm}|p{2.7cm}|p{3cm}|} 
\hline
Acquisition Time (ms) & Success Rate (\%) & Mean ebit Rate (ebits/s) & Average Total ebits & Mean Cross Correlation SNR & Mean Clock Offset Error (ps) & Mean Offset Error `Successful' cases (ps)\\
\hline
100 & 13 & 199 & 20 & 10.9 $\pm$ 1.1 & (0.6 $\pm$ 1.4)$\times 10^7$ & 68 $\pm$ 56 \\
150 & 30 & 199 & 30 & 10.5 $\pm$ 1.3 & (0.2 $\pm$ 1.2)$\times 10^7$ & 41 $\pm$ 52 \\
200 & 42 & 199 & 40 & 10.3 $\pm$ 1.5 & (0.2 $\pm$ 1.0)$\times 10^7$ & 44 $\pm$ 52 \\
250 & 67 & 199 & 50 & 10.4 $\pm$ 1.6 & (0.5 $\pm$ 10.0)$\times 10^6$ & 25 $\pm$ 60 \\
500 & 96 & 199 & 99 & 11.0 $\pm$ 1.6 & (0.4 $\pm$ 2.5)$\times 10^6$ & 80 $\pm$ 52 \\
\hline
\end{tabular}
\caption{Static simulation results: Varying acquisition time with a fixed link loss between the satellite and ground station (41 dB) and no loss for the local signal, satellite clock frequency accuracy of $3 \times 10^{-10}$, 50 ps timestamp resolution, 100 ps detector timing jitter (FWHM), and 1000 Hz dark count rate with 50\% detection efficiency per detector. The simulation includes 0 dB of loss between the photon pair source and local detectors. `Successful' cases are those whose simulated clock offset estimate is within 1 ns of the true value.}
\label{tab:static_effect_acquisition}
\end{center}
\end{table*}

\section{Summary of results}\label{sec:main_results}

The goal of this paper is to quantitatively  evaluate the performance capabilities of a satellite-based scheme to perform clock synchronization between ground stations spread across the globe using quantum resources. We have developed a software infrastructure to simulate the evolution of a satellite-network, from which we can compute the real-time quantum data communication rates between ground stations and satellites. All aspects of the network regarding synchronization capabilities can be extracted from these real-time quantum data communication rates.  To put these tools in action, we have considered a representative example, consisting in assessing the feasibility of synchronizing ground stations located at the four corners of contiguous United States (NYC - New York City, LA - Los Angeles, SEA - Seattle, ATL - Atlanta). We show through numerical simulations that a single satellite in a LEO orbit can provide 1 ns sync precision between ground clocks in such a network. The range of parameters chosen for this study are either commensurate with off-the-shelf equipment or will be available for commercial use in the near-term.
We then analyse the effect of different hardware and constellation design parameters on 3 key performance measures, the sync precision achieved, the scale of the network and the time gaps between successive connections. The important takeaways in this regard are as follows:
\begin{itemize}
    \item The sync precision achieved between clocks located in a given region on the Earth's surface is determined by the rate at which ebits can be exchanged between the ground stations and a common satellite. Thus setting an ebit rate threshold  -- minimum rate of ebit exchange between ground stations and satellites at which connection is considered established, translates to a requirement on minimum achievable precision. E.g. a threshold of 200 ebits/s translates to a ensuring a sync precision of at least 1 ns between a ground station and a satellite (see Table \ref{tab:static_effect_no_jitter_50ps_res}, 100 \% of simulations are successful with a mean error of less than 50 ps). Higher jitter, noise and/or dark counts leads to a higher threshold requirement at the same level of precision (see Table \ref{tab:static_effect_loss_100ps_jitter_50ps_res}). At a fixed level of jitter and noise, increasing the threshold translates to higher precision requirement. At the same time, since the ebit rate falls with increasing distance between the ground station and satellite, the network size falls if a higher threshold/higher precision requirement is considered. (See Table \ref{tab:scenario1results})
    \item The stability of a satellite’s onboard clock can markedly enhance the geographical extent (network scale) to which ground stations can be synchronized. E.g., in the case of a single satellite in 500 km LEO orbit, if the satellite clock has no ability to hold time, two cities at most $\approx 2000$ km apart along the equator can be connected (simultaneous links to both the cities are needed in this case (see Figure \ref{fig:duallink}). Whereas if a CSAC with the ability to hold time at 1 ns precision for around 100 s is onboard, this scale doubles to around 4000 km (and hence covering the contiguous US). See Figures \ref{fig:ebitdistance} and \ref{fig:shadows}. 
    \item Satellites in higher orbits give smaller ebit rates but longer connectivity and larger network coverage areas. This means there is a non-trivial trade-off between precision and coverage areas (see Figures \ref{fig:ebitdistance} and \ref{fig:connectiontime}). Our results show that for a 1 ns precision requirement, MEO satellites with smaller onboard clock stability can provide better connectivity for the contiguous US network, compared to LEO satellites with more stable clocks (compare Tables \ref{tab:scenario1results} and \ref{tab:MEOresults}). This also translates to an increase in the coverage area when moving from a LEO to a MEO constellation, for the same choice of onboard hardware parameters.
    \item Optimal orientation of the orbit leads to increase in network scale (e.g., a tilted orbit at $-50^\circ$ angle with the Earth's axis of rotation lets us cover the continental US whereas a non-tilted polar orbit fails to do so). See Figure \ref{fig:tilt_effect}. 
    \item Adding more satellites in an orbit (up to a threshold number) gives higher connected time percentages by reducing the length of short gaps between connections, and larger number of orbits reduces the longer time gaps between satellite and ground station connections (see Figures \ref{fig:NYC_LA_sync_uplinks} and \ref{fig:nsateffect}). Since, inter-satellite links are not considered in this work, two ground stations can only be connected if they are in view of a common satellite. It should be thus be noted that adding satellites to the constellation cannot enhance the scale of the network.
\end{itemize}
Finally, in Section \ref{subsec:static_simulations}, by using static simulations we justify the use of ebit rate thresholds as a proxy for sync precision and show the relationship between these quantities. We also study the effect of loss, dark counts and jitter on the sync precision. We further establish that the acquisition times required to perform the sync at sub-nanosecond levels is around 250 ms. An important implication of this short acquisition time is that since relativistic effects produce less than a nanosecond of relative clock drift per second between the satellite and ground station \cite{Zhu_GPS_relativity}, they can be safely ignored for the QCS protocol, if the sync requirements are at the nanosecond level.     
\section{Discussion and future work} \label{sec:Conclusions}
This paper contains a study of the feasibility and performance capabilities of a quantum clock synchronization protocol to distribute time across the globe using networks of satellites equipped with quantum resources. The protocol comes with the added advantage that  it can sync clocks independent of ranging (i.e., knowledge of distances) and hence is  more robust  to trajectory inaccuracies and delay attacks --- in addition to the extra layer of \emph{quantum security} associated with the polarization state entanglement.  We find that  large-area networks of synchronized clocks can be established with modest resources of a  handful of quantum-enabled satellites with modest equipment onboard, with expected sync precision of a few nanoseconds going up to tens of picoseconds --- higher than the GPS.  In particular, these satellites  do not need to have inter-satellite communication capabilities, and  need only to have relatively stable clocks, satellite-ground (uplink and downlink) quantum communication capabilities, and reasonable quantum hardware (entangled photon-pair sources and photodetectors). These ingredients are  common to the requirements of current space missions which seek to establish quantum communication links (e.g., long-distance quantum key distribution). Hence, our clock synchronization protocol adds a extra layer of utility to quantum technologies in the space domain. 

A takeaway of our analysis is that, by adding  more functionality to the satellites --- particularly, inter-satellite quantum communication capabilities --- it is plausible  that major cities across the globe can share a quantum-secure, highly-synchronized common time, using a modest amount of resources which could also have other functionality.  The advantages of such a ``clock network in the sky'' do not come for free, as this is a configuration that will require more resources. At the same time, this remains a feasible concept, since the inter-satellite communication occurs essentially in vacuum, involving minimum losses. The limiting factor for such configuration would still be satellite-ground station communication. The detailed study of the capabilities of such a network is the goal of future work.

The present work also contains several limitations.  We have used the quantum communication rates as a proxy for sync performance, and considered  the effects of a possible constant relative velocity between the ground stations and satellites. In a real situation, however, these  relative velocities are not constant and, as discussed in sections \ref{subsec:extraction} and \ref{sec:simulations}, to quantify the effects they have on the sync precision one needs a  real time calculation of correlation functions. Although such calculation goes beyond the scope of this work, our analysis shows that the time needed for a ground station to synchronize with a satellite (what we have called the acquisition time) is sufficiently short for the relative velocities to be well approximated by a constant value. Although the concrete value of the acquisition time depends on the losses and noise levels (other hardware parameters remaining fixed), our calculations show that typical values are not more than a few hundred milliseconds. This interval is short enough for the relative velocity between satellite and ground stations to remain constant.

Similarly, we have not included the impact of relativistic effect in this work. One could have the impression that, since relativist effects play an important role in other satellite based synchronization systems, and in particular for the GPS, they should play an even more important role in the protocol we study in this paper, since it aims at a better sync precision. Again, the key observation is that the acquisition time required to establish synchronization is estimated to be not more than a fraction of a second, and such interval is not long enough for relativistic effects to build up significantly \cite{Zhu_GPS_relativity}. Conversely, if synchronization can be established fast enough in realistic setting, the quantum clock synchronization network could be thought as a precise measurement device for relativistic effects around the Earth. Many of the leading and higher order relativistic effects could be in  range of measurement through a near-term implementation of the ideas proposed here. This in turn would also provide a way to measure the effects of weak gravity on quantum systems, complementing dedicated studies \cite{Qoptics_space_exp,Anthony_HOM_grav,Bell_test_weak_grav,Einstein_eq_Qopt,neutron_gravAbele2012,SAGE}. These and other extensions ---such as the use of realistic orbits, aspects of the extra security layer added by the entanglement in polarization of photons, etc.---will be the focus of future work.

\acknowledgments
This work was partially supported by the US SBIR/STTR program under Air Force Research Laboratory contract FA9453-20-P-0100 and NASA contract 80NSSC20C0395. JET acknowledges additional support from Xairos Systems, Inc. I.A. is supported by the NSF grants PHY-2110273, from Xairos Systems Inc., and from the Hearne Institute for Theoretical Physics at Louisiana State University.

\bibliography{report} 
\bibliographystyle{apsrev4-1}

\end{document}